\documentclass[a4paper,12pt]{article}
\pdfoutput=1
\usepackage[margin=1in]{geometry}
\usepackage{graphicx,xcolor}
\usepackage{mathtools,braket,blkarray}
\allowdisplaybreaks
\usepackage{amssymb,amsfonts,bbm,relsize}
\usepackage[width=0.9\textwidth,footnotesize]{caption}
\usepackage[font=scriptsize]{subcaption}
\usepackage{cite,hyperref,url}
\hypersetup{pageanchor=false}
\usepackage{booktabs,multirow,makecell}
\usepackage[utf8]{inputenc}

\newcommand{\overbar}[1]{\mkern 1.5mu\overline{\mkern-1.5mu#1\mkern-1.5mu}\mkern 1.5mu}
\newcommand{\pmatr}[1]{\begin{pmatrix} #1 \end{pmatrix}}
\newcommand{\simlt}{~\mbox{\smaller\(\lesssim\)}~}
\newcommand{\simgt}{~\mbox{\smaller\(\gtrsim\)}~}
\newcommand{\hvev}[2]{\braket{H_{45}^{#1}}_{#2}}

\begin{document}

\begin{titlepage}
\begin{center}
{\bf\Large A natural \texorpdfstring{\boldmath{$S_4 \times SO(10)$}}{S4xSO10} model of flavour } \\[12mm]
Fredrik~Bj\"{o}rkeroth$^{\star}$%
\footnote{E-mail: {\tt f.bjorkeroth@soton.ac.uk}},
Francisco~J.~de~Anda$^{\dagger}$%
\footnote{E-mail: \texttt{franciscojosedea@gmail.com}},
Stephen~F.~King$^{\star}$%
\footnote{E-mail: \texttt{king@soton.ac.uk}},
Elena~Perdomo$^{\star}$%
\footnote{E-mail: \texttt{e.perdomo-mendez@soton.ac.uk}}
\\[-2mm]

\end{center}
\vspace*{0.50cm}
\centerline{$^{\star}$ \it
School of Physics and Astronomy, University of Southampton,}
\centerline{\it
SO17 1BJ Southampton, United Kingdom }
\vspace*{0.2cm}
\centerline{$^{\dagger}$ \it
Tepatitl{\'a}n's Institute for Theoretical Studies, C.P. 47600, Jalisco, M{\'e}xico}
\vspace*{1.20cm}

\begin{abstract}
{\noindent
We propose a natural $ S_4 \times SO(10) $ supersymmetric grand unified theory of flavour with an auxiliary $\mathbb{Z}_4^2 \times \mathbb{Z}_4^R$ symmetry, based on small Higgs representations (nothing larger than an adjoint) and hence a type-I seesaw mechanism.
The Yukawa structure of all fermions is determined by the hierarchical vacuum expectation values of three $ S_4 $ triplet flavons, with CSD3 vacuum alignments, where up-type quarks and neutrinos couple to one Higgs \textbf{10}, and the down-type quarks and charged leptons couple to a second Higgs \textbf{10}. 
The Yukawa matrices are obtained from sums of low-rank matrices, where each matrix in the sum naturally accounts for the mass of a particular family, as in sequential dominance in the neutrino sector, which predicts a normal neutrino mass hierarchy.
The model accurately fits all available quark and lepton data, with predictions for the leptonic $CP$ phase in 95\% credible intervals given by 
$ 281^\circ < \delta^\ell < 308^\circ $ and
$ 225^\circ < \delta^\ell < 253^\circ $.
The model reduces to the MSSM, with the two Higgs doublets emerging from the two Higgs \textbf{10}s without mixing, and we demonstrate how a $\mu$ term of $\mathcal{O}$(TeV) can be realised, as well as doublet-triplet splitting, with Planck scale operators controlled by symmetry, leading to acceptable proton decay. 
}
\end{abstract}
\end{titlepage}

\section{Introduction}

The flavour problem, the origin of the three families of quarks and leptons with their observed pattern of masses and mixing, remains one of the longest-standing and deepest mysteries left unanswered by the Standard Model (SM) \cite{Olive:2016xmw}.
Among its features are very hierarchical charged fermion masses, with the up-type quark mass hierarchy 
$m_u\ll m_c\ll m_t$ being stronger than for the down-type quark masses
$m_d\ll m_s \ll m_b$ 
which resemble more closely the charged lepton masses
$m_e\ll  m_{\mu}\ll  m_{\tau}$. 
The lightest charged fermion is the electron, with
$m_e\sim 0.5$ MeV,
while the heaviest (third family) masses satisfy $m_{\tau} \sim m_b \ll m_t$ at high energies.
Quark mixing is small and hierarchical: 
$V_{ub}\sim \lambda^3$,  
$V_{cb}\sim \lambda^2$, 
$V_{us}\sim \lambda$,
where $\lambda \approx 0.225$ \cite{Olive:2016xmw}. 

The discovery of neutrino mass and mixing \cite{Fukuda:1998mi}, elucidated over the past twenty years \cite{nobel}, makes the flavour problem more acute \cite{King:2017guk}, but also provides new features, namely small neutrino masses, and large lepton mixing resembling tri-bimaximal (TB) mixing, but with non-zero reactor angle:
$U_{e2}\approx U_{\mu 2}\approx U_{\tau 2}\approx \frac{1}{\sqrt{3}}$, 
$U_{\mu 3}\approx U_{\tau 3}\approx \frac{1}{\sqrt{2}}$,
$U_{e 3}\approx \frac{\lambda}{\sqrt{2}}$ \cite{King:2012vj}.
The origin, nature and ordering of the neutrino masses remain open questions, but cosmology suggests that all neutrino masses must be below about 100 meV \cite{Ade:2015xua}, making them by far the lightest (known) fermions in nature.
There are hints of $CP$ violation in the lepton sector \cite{Esteban:2016qun} but so far it has not been established, unlike in the quark sector where the $CP$ phase is well measured \cite{Olive:2016xmw}.

One of the most attractive and popular possibilities for generating small neutrino masses is the type-I seesaw mechanism involving three right-handed (RH) neutrinos \cite{Minkowski:1977sc}.
A natural way to obtain large lepton mixing and normal neutrino hierarchy within type-I seesaw is to assume the sequential dominance (SD) of RH neutrinos \cite{King:1998jw}.
SD postulates three RH neutrinos, where one of them, usually the heaviest one, is almost decoupled from the seesaw mechanism, and is responsible for the lightest physical neutrino mass $m_1$.
Of the remaining two, one gives the dominant seesaw contribution and is mainly responsible for the (heaviest) atmospheric neutrino mass $m_3$ and mixing, while the other gives a subdominant contribution, responsible for the (second-heaviest) solar neutrino mass $m_2$ and mixing. 
SD therefore predicts $m_1\ll m_2 \ll m_3\sim 50$ meV.%
The magnitude of atmospheric and solar mixing is determined by ratios of Yukawa couplings, which can easily be large, while the reactor mixing is typically $ U_{e3} \simlt \mathcal{O}(m_2/m_3) \approx 0.17 $.
This successful prediction was made over a decade before the reactor angle was measured \cite{reactorangle}.

To obtain precise predictions for mixing one can go further and impose constraints on the Yukawa couplings, as in constrained sequential dominance (CSD) \cite{King:2005bj,King:2013iva,King:2013xba,Bjorkeroth:2014vha,King:2015dvf}.
A particularly successful scheme is known as CSD3 \cite{King:2013iva,Bjorkeroth:2014vha,King:2015dvf,King:2016yvg,Ballett:2016yod} where the neutrino Yukawa matrix is controlled by particular vacuum expectation values (VEVs) of three triplet flavon fields, $ \braket{\phi_i} $, as discussed later.
The flavon vacuum alignments are fixed by a superpotential which we do not specify here, but may be enforced by an $S_4$ symmetry, as discussed in \cite{King:2016yvg}. 
After implementing the seesaw mechanism, the above flavons yield a light effective left-handed (LH) Majorana neutrino mass matrix,
\begin{equation}
	m^\nu = \mu_1 Y_{11} + \mu_2 Y_{22} + \mu_3 Y_{33},
\label{eq:mnu}
\end{equation}
where $Y_{ij}\sim \braket{\phi_i}\braket{\phi_j}^T$, up to $S_4$ Clebsch-Gordan (CG) factors.
Each of the matrices $Y_{ii}$ is quadratic in $ \braket{\phi_i} $ and therefore has rank 1.
The SD condition implies that $\mu_2 > \mu_1$ and hence maximal atmospheric mixing is controlled by $Y_{22}$, solar mixing is controlled by $Y_{11}$, while $Y_{33}$ plays no important role in neutrino physics due to the smallness of $\mu_3$, which implies that $m_1$ is similarly small.
In the limit $ \mu_3 = m_1 = 0 $, this effectively reduces to a two RH neutrino model, with only two real parameters $\mu_1$ and $\mu_2$. The resulting model is known as Littlest Seesaw (LS)~\cite{King:2015dvf,King:2016yvg,Ballett:2016yod}.%
\footnote{
	The LS neutrino mass matrix in Eq.~\ref{eq:mnu} annihilates the first column of the TB mixing matrix, leading to TM$_1$ mixing with a fixed reactor angle.
}
It is remarkable that, with $ \mu_{1,2} $ fitted to the neutrino masses, the entire PMNS matrix is then uniquely determined with no free parameters, giving predictions for mixing angles and the $CP$ phase in agreement with current data. 

In order to extend the above ideas into a unified model of flavour, the CSD3 model with two RH neutrinos may be embedded into an $ A_4 \times SU(5) $ Grand Unified Theory (GUT) \cite{Bjorkeroth:2015ora}.
However, from a theoretical point of view, the $SO(10)$ GUT is preferred, since it predicts three RH neutrinos and makes neutrino mass inevitable.
This motivated the $ \Delta_{27} \times SO(10) $ model \cite{Bjorkeroth:2015uou}, where the mass matrices for all quarks and leptons had the same universal form as the neutrino mass matrix in Eq.~\ref{eq:mnu}, but with different coefficients multiplying each rank-1 matrix. 
While successful and rather complete, there were several unsatisfactory features of the $\Delta_{27}\times SO(10)$ model. It was quite a complicated model involving a large number of fields.
Furthemore, $\Delta_{27}$ cannot enforce the desired CSD3 vacuum alignments by symmetry alone, unlike $S_4$, which therefore seems to be preferred by CSD3. 
Finally, a universal mass matrix structure, which seems quite appealing at first sight, led to problems in the quark sector, which were fixed by adding an extra non-universal term in the up-type quark sector, together with some degree of fine-tuning between matrix coefficients in order to obtain the correct quark masses and mixing angles. 

In the present paper we propose a natural $S_4\times SO(10)$ grand unified theory of flavour in which the CSD3 model of neutrinos is embedded.
Our guiding principles are 
firstly simplicity, involving the fewest number of low-dimensional fields, 
secondly naturalness, and 
thirdly completeness, in particular addressing the doublet-triplet splitting problem.
What does natural mean? 
For us it means that we have a qualitative explanation of charged fermion mass and mixing hierarchies, as for neutrino mass and mixing, with all dimensionless parameters $ \mathcal{O}(1) $, and in particular that the Yukawa matrices are obtained from sums of low-rank matrices, as in Eq.~\ref{eq:mnu}, where each matrix in the sum naturally accounts for the mass of a particular family, analogous to SD in the neutrino sector. 
This qualitative picture of ``universal sequential dominance'' is underpinned by a detailed quantitative fit of the fermion spectrum. 

In order to achieve this, we shall introduce two Higgs \textbf{10}s, $H^u_{10}$ and $H^d_{10}$, which will give rise, at low energy, to the minimal supersymmetric standard model (MSSM) Higgs doublets, $h_u$ and $h_d$, respectively, with no appreciable Higgs mixing effects.
Neutrinos and up-type quarks, which couple to $H^u_{10}$, have Yukawa matrices with a universal structure as in Eq.~\ref{eq:mnu}, dictated by CSD3.
The charged leptons and down-type quarks, which couple to $H^d_{10}$, have Yukawa matrices with a different universal structure where $Y_{11}$ is replaced by $Y_{12} \sim \braket{\phi_1}\braket{\phi_2}^T$.
Then quark mixing originates primarily in the down-type quark sector, with the down and strange quark masses successfully realised by having a zero entry in the (1,1) element of the down-type quark Yukawa matrix $ Y^d $, as in the GST approach \cite{Gatto:1968ss}, with a milder hierarchy among down-type quarks as compared to up-type quarks.

The model accurately fits all available quark and lepton data, and predicts 
the leptonic $CP$ phase in 95\% credible intervals given by 
$ 281^\circ < \delta^\ell < 308^\circ $ and
$ 225^\circ < \delta^\ell < 253^\circ $, with 
a significant deviation from maximal $CP$ violation.
Since quark mixing dominantly originates from $ Y^d $, analytical estimates for the quark mixing angles are given, revealing some tension in the predicted observables, which can be ameliorated by assuming rather large SUSY threshold corrections.
A hierarchy in the flavon VEVs fixes the scales of all but one parameter, with all dimensionless couplings in the renormalisable theory naturally $\mathcal{O}$(1).
The model reduces to the MSSM, and we demonstrate how a $\mu$ term of $\mathcal{O}$(TeV) can be realised, as well as doublet-triplet splitting, with Planck scale proton decay operators suppressed. 
In order to achieve the above we also require an auxiliary $\mathbb{Z}_4^2 $ and $\mathbb{Z}_4^R$ symmetry and a spectrum of messenger fields.

We would like to emphasise that the model presented here is very different from earlier models based on $S_4\times SO(10)$
\cite{HagedornLindnerMohapatra,DuttaMimuraMohapatra,BhupalDevDuttaMohapatraSeverson,Patel} 
(see also \cite{LeeMohapatra,CaiYu,BhupalDevMohapatraSeverson}).%
\footnote{
	Previous works on $ SO(10) $ models with non-Abelian discrete flavour symmetries are found in \cite{so10}, and further flavoured GUTs can be found in \cite{nonAbelian}. More recently, a generalised approach to flavour symmetries in $ SO(10) $ is considered in \cite{Ivanov:2015xss}.
}
Firstly, the full symmetry is different, since we invoke an extra $\mathbb{Z}_4^2\times \mathbb{Z}_4^R$ symmetry, while earlier works use a $\mathbb{Z}_n$ \cite{DuttaMimuraMohapatra,BhupalDevDuttaMohapatraSeverson,Patel}.
Furthermore, we only allow small Higgs representations \textbf{10} (fundamental), 
\textbf{16} (spinor) and \textbf{45} (adjoint) and do not allow the large Higgs representations such as the 
$\mathbf{\overbar{126}}$ and \textbf{120}  which are used in the other approaches.
As a consequence our neutrino masses follow from a type-I seesaw mechanism, rather than a type-II seesaw employed in other papers. 
In further contrast, we do not allow Higgs mixing: the MSSM Higgs doublets $h_u$ and $h_d$ emerge directly from $H^u_{10}$ and $H^d_{10}$, respectively, whereas in \cite{HagedornLindnerMohapatra,DuttaMimuraMohapatra,BhupalDevDuttaMohapatraSeverson,Patel} they arise as unconconstrained linear combinations of doublets contained in 10- and 126-dimensional Higgs fields. 
In addition we consider doublet-triplet splitting.
These features are largely absent from earlier works.

Another important difference is that we have used the CSD3 vacuum alignments in \cite{King:2016yvg}, whereas the vacuum aligments used in most previous works were geared towards TB mixing, and do not naturally provide a large reactor angle.
For instance, in \cite{DuttaMimuraMohapatra} they predict $ U_{e3} \sim 0.05 $, which is now excluded.
Indeed our model is motivated by the success of CSD3 and LS in the neutrino sector, although the LS predictions will be slightly modified due to charged lepton mixing corrections which are necessarily present due to the relation between charged leptons and down-type quarks.
We should also mention that the first attempts to unify the three families as $\psi \sim (3,16)$ with Yukawa matrices resulting from flavons with particular vacuum alignments were in \cite{King:2001uz}.

The remainder of the paper is laid out as follows: in Section \ref{sec:model} we describe the symmetries and field content of the model, the diagrams which yield the fermion mass matrices, and discuss proton decay. 
In Section \ref{sec:massmatrices} we define the mass and Yukawa matrices, derive analytical estimates for quark mixing parameters, and perform a detailed numerical fit to data. 
Section \ref{sec:conclusion} concludes. 
In appendices we describe the mechanism giving doublet-triplet splitting and a $\mu$ term, and the correct treatment of the third family Yukawa couplings, as well as a detailed derivation of the mass matrices.

\section{The model}
\label{sec:model}
We present a model with quarks and leptons unified in a single $\psi $ in the $(3',16)$ representation of $S_4\times SO(10)$, and with $H^{u,d}_{10}$ in $(1,10)$ and $\phi_i$ in $(3',1)$ representations.
The idea is that the up-type quark Yukawa matrix $ Y^u $ and neutrino Yukawa matrix $ Y^\nu $ arise from effective terms like
\begin{equation}
	H^u_{10} (\psi \phi_1)(\psi \phi_1)+ 
	H^u_{10} (\psi \phi_2)(\psi \phi_2) + 
	H^u_{10} (\psi \phi_3)(\psi \phi_3),
 \label{eq:introup}
\end{equation}
where the group contraction in each bracket is into an $S_4$ singlet.
These nonrenormalisable operators will have denominator scales of order $M_\mathrm{GUT}$, determined by the VEVs of additional Higgs adjoint \textbf{45}s, leading to various CG factors.
This leads to the Yukawa matrices $ Y^u $ and $ Y^\nu $ being a sum of rank-1 matrices as in Eq.~\ref{eq:mnu}, with independent coefficients multiplying each rank-1 matrix, where $Y_{ij} \sim \braket{\phi_i}\braket{\phi_j}^T$, up to $S_4$ CG factors.
In the present model we assume the CSD3 flavon vacuum alignments \cite{King:2016yvg},
\begin{equation}
	\braket{\phi_1} = v_1 \pmatr{1\\3\\-1}, \quad 
	\braket{\phi_2} = v_2 \pmatr{0\\1\\-1}, \quad
	\braket{\phi_3} = v_3 \pmatr{0\\1\\0},
\label{eq:flavons}
\end{equation}
with VEVs driven to scales with the hierarchy%
\footnote{
	In the full model we shall not provide an explanation for this hierarchy of VEVs, nor shall we repeat the vacuum alignment superpotential responsible for the alignments in Eq.~\ref{eq:flavons}, which is discussed in \cite{King:2016yvg}. We note that the alignments $ \braket{\phi_1} $, $ \braket{\phi_2} $ preserve the $ SU $ generator of $ S_4 $.
}
\begin{equation}
	v_1 \ll v_2 \ll v_3 \sim M_\mathrm{GUT},
\label{eq:flavonhierarchy}
\end{equation}
so that each rank-1 matrix in the sum contributes dominantly to a particular family, giving a rather natural understanding of the hierarchical Yukawa couplings
$y_u\sim v_1^2/M_\mathrm{GUT}^2$, 
$y_c\sim v_2^2/M_\mathrm{GUT}^2$, 
$y_t\sim v_3^2/M_\mathrm{GUT}^2$,
and similarly for the neutrino Yukawa couplings.
Since the expansion breaks down for the third family, in the complete model we shall find a renormalisable explanation of the third-family Yukawa couplings.
The RH neutrino Majorana mass matrix will also have the same universal form, leading to the seesaw mass matrix as in Eq.~\ref{eq:mnu}.

The down-type quark Yukawa matrix $ Y^d $ and charged lepton Yukawa matrix $ Y^e $ arise from terms like
\begin{equation}
	H^d_{10} (\psi \phi_1)(\psi \phi_2) + 
	H^d_{10} (\psi \phi_2)(\psi \phi_2) + 
	H^d_{10} (\psi \phi_3)(\psi \phi_3) ,
\label{eq:introdown}
\end{equation}
introducing a mixed term involving $\phi_1$ and $\phi_2$, leading to a new rank-2 Yukawa structure $Y_{12}\sim \braket{\phi_1}\braket{\phi_2}^T$.
This leads to the Yukawa matrices $ Y^d $ and $ Y^e $ having $ Y_{11} $ replaced by $ Y_{12} $, which has two consequences: it enforces a zero in the (1,1) element of $ Y^d $, giving the GST relation for the Cabibbo angle, i.e. $ \theta_{12}^q \approx \sqrt{y_d/y_s} $, and also leads to a milder hierarchy in the down and charged lepton sectors. 
Both features are welcome. 

We need additional symmetries and fields to ensure the above structures, provide renormalisable third family Yukawa couplings, give the desired CG relations to distinguish down-type quarks from charged leptons, achieve doublet-triplet splitting, and obtain the MSSM Higgs doublets $h_u$ and $h_d$ from $H^u_{10}$ and $H^d_{10}$, respectively.
Two $\mathbb{Z}_4$ shaping symmetries help to forbid any mixed flavon Yukawa terms. 
We also assume a discrete $ R $ symmetry $ \mathbb{Z}_4^R $, under which the superpotential has total charge two, and which is broken at the GUT scale by the $H_{45}^{B-L}$ VEV to $ \mathbb{Z}_2^R $, the usual $ R $ (or matter) parity in the MSSM, ensuring a stable lightest supersymmetric particle (LSP).
It also controls the $\mu$ term and helps ensure that only two light Higgs doublets (and no Higgs triplets) are present in the effective MSSM.
$ \mathbb{Z}_4^R $ is the smallest $ R $ symmetry that can achieve the above, and is specially motivated within $SO(10)$ \cite{Lee:2010gv}. 
We shall also assume a spontaneously broken 
canonical
$ CP $ symmetry at the high scale.

The full superfield content of the model is given in Table \ref{tab:funfields}. 
It contains the following: a ``matter'' superfield $ \psi $ containing all known SM fermions, three triplet flavons $ \phi $ which acquire CSD3 vacuum alignments, two Higgs \textbf{10}s containing one each of the electroweak-scale Higgs $ SU(2) $ doublets,%
\footnote{
	We assume that the MSSM Higgs doublets $ h_u, h_d $ lie completely inside, respectively, the $ SO(10) $ multiplets $ H^u_{10}, H^d_{10} $. 
	This is justified in Appendix~\ref{sec:app:DTsplitting}.
}
a spinor $ H_{\overbar{16}} $ which breaks $ SO(10) $ (and gives masses to the RH neutrinos), as well as several Higgs adjoints. 
The $ \chi $ superfields are messengers that are integrated out below the GUT scale, and are given GUT-scale masses by the VEV of $ H_{45}^Z $.

\begin{table}[ht]
\begin{subtable}[b]{0.5\textwidth}
\centering
	\begin{tabular}{ l c@{\hskip 5pt} c c c c}
	\toprule
	\multirow{2}{*}{\rule{0pt}{4ex}Field}	& \multicolumn{5}{c}{Representation} \\
	\cline{2-6}
	\rule{0pt}{3ex}
	& $S_4$ & $SO(10)$ &$\mathbb{Z}_{4}$ & $\mathbb{Z}_{4}$ &  $\mathbb{Z}_{4}^R$ \\
	\midrule
	$\psi$ 		 & $3^\prime$ & 16 & 1 & 1 &1 \\
	\rule{0pt}{3ex}%
	$H_{10}^{u}$ & 1 & 10 & 0 & 2 & 0 \\
	$H_{10}^{d}$ & 1 & 10 & 2 & 0 & 0\\
	$H_{\overbar{16}}$ & 1 & $\overbar{16}$ & 2 & 1 & 0\\
	$H_{16}$ & 1 & 16 & 1 & 2 & 0\\
	$H_{45}^{X,Y}$ & 1 & 45 & 2 & 1 & 0 \\
	$H_{45}^Z$ & 1 & 45 & 1 & 2 & 0 \\
	$H_{45}^{B-L}$ & 1 & 45 & 2 & 2 & 2\\
	\rule{0pt}{3ex}%
	$\phi_1$ & $3^\prime$ &1 & 0 & 0 & 0\\
	$\phi_2$ & $3^\prime$ &1 & 2 & 0 & 0\\
	$\phi_3$ & $3^\prime$ &1 & 0 & 2 & 0\\
	\bottomrule
	\end{tabular}
\caption{Matter, Higgs and flavon superfields.}
\end{subtable}
\begin{subtable}[b]{0.5\textwidth}
\centering
	\begin{tabular}{ c c@{\hskip 5pt} c c c c}
	\toprule
	\multirow{2}{*}{\rule{0pt}{4ex}Field}	& \multicolumn{5}{c}{Representation} \\
	\cline{2-6}
	\rule{0pt}{3ex}
	& $S_4$ & $SO(10)$ &$\mathbb{Z}_{4}$ & $\mathbb{Z}_{4}$ &  $\mathbb{Z}_{4}^R$ \\

	\midrule
	$\overbar{\chi}_1$ & 1 &$\overbar{16}$	& 3 & 3 & 1\\
	${\chi}_1$ & 1 & 16 					& 0 & 3 & 1 \\
	$\overbar{\chi}_2$ & 1 &$\overbar{16}$	& 1 & 3 & 1 \\
	${\chi}_2$ & 1 & 16 					& 2 & 3 & 1 \\
	$\overbar{\chi}_3$ & 1 & $\overbar{16}$	& 3 & 1 & 1 \\
	$\chi_3$ & 1 & 16 						& 0 & 1 & 1\\
	${\chi}_3^\prime$ & 1 & 16 				& 3 & 2 & 1\\
	${\chi}_2^\prime$ & 1 & 16 				& 1 & 0 & 1\\
	\rule{0pt}{3ex}%
	$ \rho $ & 1 & 1 & 2 & 2 & 1\\
	\bottomrule
	\end{tabular}
\caption{Messenger superfields.}
\end{subtable}
\caption{Field content giving the Yukawa superpotential in Eq.~\ref{eq:WYrenorm}.}
\label{tab:funfields}
\end{table}

At the GUT scale, the renormalisable Yukawa superpotential is given by
\begin{equation}
\begin{split}
	W_Y^\mathrm{(GUT)} &= 
		\psi \phi_a \overbar{\chi}_a 
		+ \overbar{\chi}_a \chi_a H_{45}^Z 
		+ \chi_a \chi_a H^u_{10} 
		+ \rho \chi_3 H_{\overbar{16}} + M_\rho \rho \rho
	\\ &\quad 
	+ 
		\overbar{\chi}_b \chi_b^\prime \left(H_{45}^X + H_{45}^Y\right) 
		+ \chi^\prime_b \chi^\prime_b H_{10}^d 
		+ \chi_1\chi_2H^d_{10} ,
\label{eq:WYrenorm}
\end{split}
\end{equation}
where we sum over indices $ a = 1,2,3 $ and $ b = 2,3 $, and have suppressed $ \mathcal{O}(1) $ coefficients $ \lambda $ that multiply each term.
Furthermore, there are several crucial terms that appear suppressed by one Planck mass $ M_P $. These are
\begin{equation}
	W_Y^\mathrm{(Planck)} = 
		\frac{\chi_a \chi_a H_{\overbar{16}} H_{\overbar{16}}}{M_P}
		+ \frac{\psi \psi \phi_3 H^d_{10}}{M_P} ,
\label{eq:WYplanck}
\end{equation}
where $ a = 1,2,3 $. 
The first term couples $ H_{\overbar{16}} $ to fermions via the messengers. The second is allowed by the symmetries and will be shown to contribute at the order of the smallest GUT-scale terms to the fermion Yukawa matrices, and thus cannot be ignored.

The adjoint Higgs superfields acquire VEVs at the GUT scale, i.e. $ \braket{H_{45}^{k}} \sim M_{\mathrm{GUT}} $, which are generally complex. 
$ H_{45}^{X,Y,Z} $ gain arbitrary (SM-preserving) VEVs, providing CG factors which separate the quark and lepton masses. 
$H_{45}^{B-L}$ gains a VEV in the direction that preserves $ B-L $, generating GUT-scale masses for Higgs triplets via the Dimopoulos-Wilczek (DW) mechanism \cite{DW}. Our implementation of the DW mechanism is described in Appendix~\ref{sec:app:DTsplitting}.

The VEVs of $ \phi_1 $ and $ \phi_2 $ are assumed to acquire VEVs well below the GUT scale, i.e. $ \braket{\phi_{1,2}} \ll M_\mathrm{GUT} $, while $ \braket{\phi_3} \sim M_\mathrm{GUT} $, which is therefore also the scale at which the flavour symmetry is broken, along with $CP$. 
We note that no residual $ CP $ symmetry remains at low scales, 
but $ CP $ does play a role in fixing phases in the mass matrices.
As $ \braket{\phi_3} $ is near the messenger scale, the process of integrating out messengers $ \chi_3, \overbar{\chi}_3 $ is not trivial. 
The correct procedure and the consequences of having a flavon VEV near $ M_\mathrm{GUT} $ are discussed in detail in Appendix~\ref{sec:app:third}, where we verify also that the third family Yukawa couplings are renormalisable at the electroweak scale. 

\begin{figure}[ht]
\centering
	\includegraphics[scale=0.7]{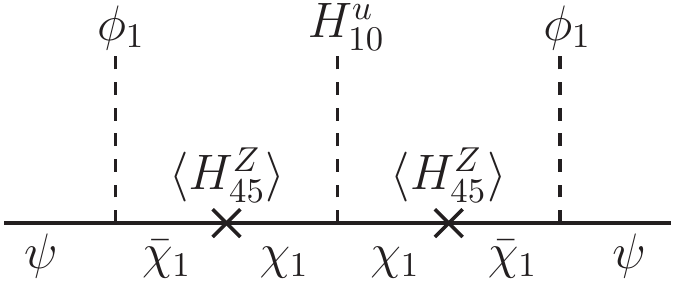}
\hspace*{1ex}
	\includegraphics[scale=0.7]{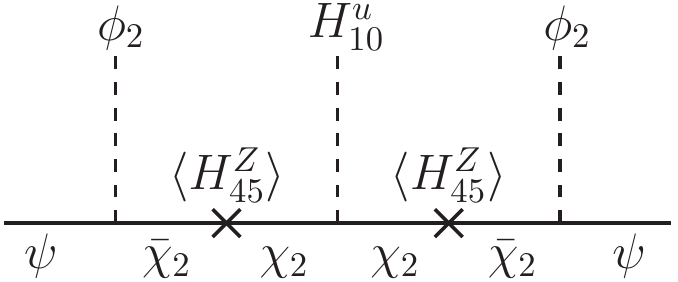}
\hspace*{1ex}
	\includegraphics[scale=0.7]{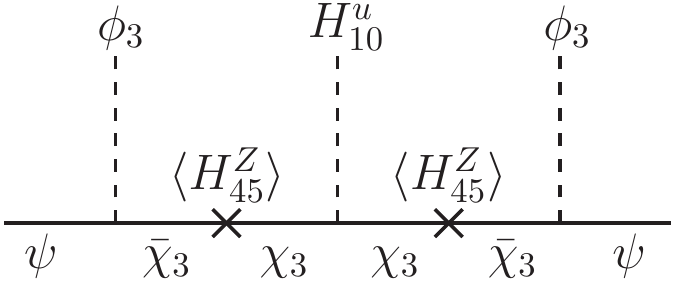}
\caption{Diagrams coupling $ \psi $ to $ H^u_{10} $. When flavons acquire VEVs, these give the up-type quark and Dirac neutrino Yukawa matrices.}
\label{fig:hu}
\end{figure}

\begin{figure}[ht]
\centering
	\includegraphics[scale=0.7]{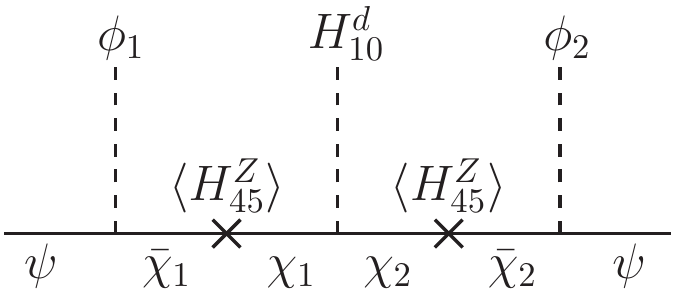}
\hspace*{1ex}
	\includegraphics[scale=0.7]{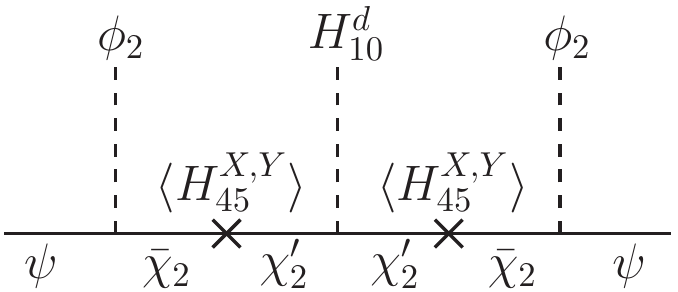}
\hspace*{1ex}
	\includegraphics[scale=0.7]{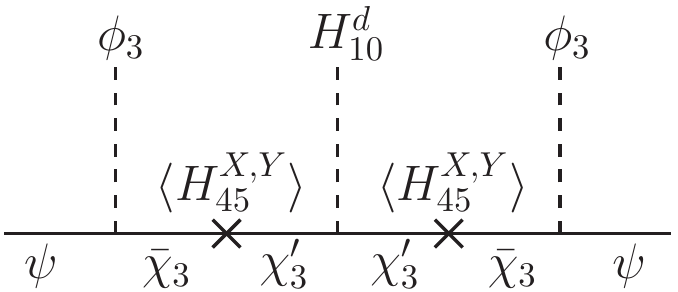}
\caption{Diagrams coupling $ \psi $ to $ H^d_{10} $. When flavons acquire VEVs, these give the down-type quark and charged lepton Yukawa matrices.}
\label{fig:hd}
\end{figure}

\begin{figure}[ht]
\centering
	\includegraphics[scale=0.7]{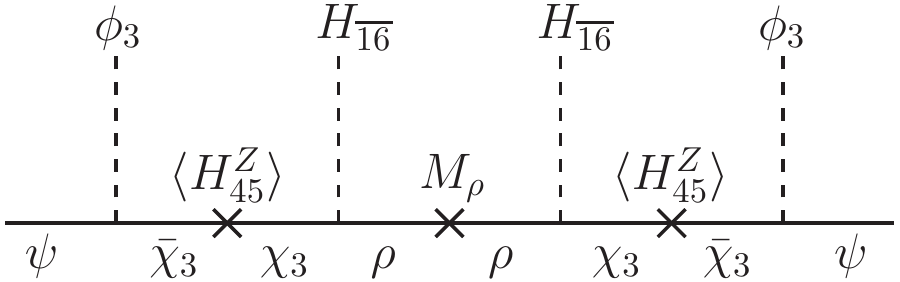}
\hspace*{2ex}
	\includegraphics[scale=0.7]{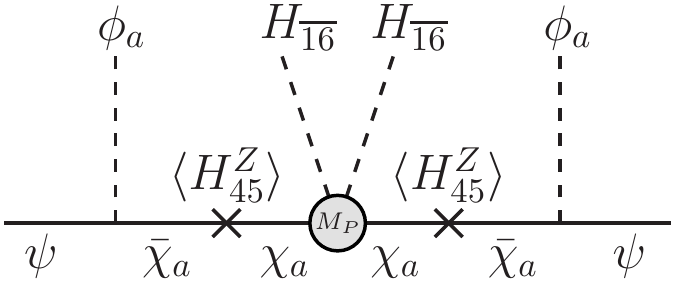}
\caption{Diagrams coupling $ \psi $ to $ H_{\overbar{16}} $. One copy of the right diagram may be drawn for each of $ a = 1,2,3 $, although for $ a = 3 $, its contribution is negligible compared to the left diagram. When flavons acquire VEVs, these give the RH neutrino mass matrix.}
\label{fig:h16}
\end{figure}

The diagrams giving the mass and Yukawa matrices are drawn in Figs.~\ref{fig:hu}-\ref{fig:h16}.%
\footnote{
	The diagrams were drawn with \texttt{JaxoDraw} \cite{Binosi:2003yf}.
}  
The three diagrams in Fig.~\ref{fig:hu} correspond to the ultraviolet completion of the three terms in Eq.~\ref{eq:introup}, while those in Fig.~\ref{fig:hd} are the completion of the terms in Eq.~\ref{eq:introdown}.
The diagrams ensure correct $S_4$ group theory contractions and introduce CG coefficients due to the $H_{45}^{X,Y,Z}$ VEVs. 
These diagrams are analogous to how the seesaw mechanism replaces the Weinberg operator for neutrino mass. 
Of course neutrino mass itself in this model is more subtle, since both the Dirac and RH Majorana masses arise from these diagrams.

Each diagram leads to a $ 3\times 3 $ matrix, whose internal structure is dictated by the vacuum alignment of the relevant flavon VEVs in Eq.~\ref{eq:flavons}. 
The Yukawa and mass matrices are consequently given as a sum over these matrices. 
A prominent feature is a texture zero in the (1,1) element of $ Y^d $ and $ Y^e $, which realises the GST relation for the Cabibbo angle.
The exact matrices that we fit to data are given in Section \ref{sec:massmatrices}, with a full derivation in Appendix~\ref{sec:app:fullderivation}.

Further Planck-scale operators suppressed by one power of the Planck mass $ M_P $ are forbidden by the symmetries. 
However we expect additional effective operators arising in the model, suppressed by at least two powers of the Planck mass $ M_P^2 $. 
These include terms involving all possible contractions of $ S_4 $ multiplets $ \psi $ and $ \phi_i $, which are forbidden at the renormalisable level, but allowed by the symmetries. 
The largest of these terms can be $ \mathcal{O}(M_\mathrm{GUT}^2/M_P^2) \sim 10^{-6} $. 
We will assume these contributions are negligible, but note that such corrections may pollute the texture zero in $ Y^d $.

An adjoint of $SO(10)$ can acquire a VEV aligned in the direction of any of the four $U(1)$ subgroup generators that commute with the SM, or a combination thereof.%
\footnote{
	There are four such $U(1)$ symmetries, labelled $U(1)_{X}$, $U(1)_{Y}$, $U(1)_{B-L}$, $U(1)_{T^3_R}$. $ U(1)_{X} $ arises from the breaking $ SO(10) \to SU(5) \times U(1)_X $. $ U(1)_Y $ is the SM hypercharge which arises when $ SU(5) \to SU(3)\times SU(2)_L\times U(1)_Y $. The other two $U(1)$ arise when $ SO(10) $ is broken along the Pati-Salam chain, via a $LR$-symmetric gauge group. Their generators are not linearly independent; two of them may be expressed in terms of the other two.
}
The VEVs of $ H_{45}^{X,Y,Z} $ may be written as linear combinations of these alignments.
Fermions couple to these VEVs with strengths that depend on their associated $ U(1) $ charges, which are different for quarks and leptons.

Up-type quarks and Dirac neutrinos couple to $ H_{45}^Z $ (see Fig.~\ref{fig:hu}).
As $ \braket{H_{45}^Z} $ is arbitrary, there is no hard prediction for the ratio between quark and neutrino Yukawa couplings within a family. 
However, as all flavons $ \phi_a $ couple to this VEV in the same way, flavour unification demands that the same ratio hold for all families. 
Therefore, once $ Y^u $ is determined, $ Y^\nu $ is also fixed, such that $ Y^\nu \propto Y^u $, to 
good approximation, up to an overall CG factor, with small deviations for the third family.
While the elements of $ Y^\nu $ are not accessible at low energy scales (e.g. in neutrino oscillations), neutrino Yukawa couplings may be probed by considering leptogenesis.
Since the lightest right-handed neutrino $N_1$ is too light, we anticipate that 
$ N_2 $ thermal leptogenesis will be required, where 
the second-to-lightest right-handed neutrino in our model has a mass of $ \mathcal{O}(10^{10}) $ GeV, which is in the preferred range.

Meanwhile, the down-type quarks and charged leptons couple to two adjoints $ H_{45}^X $ and $ H_{45}^Y $ (see Fig.~\ref{fig:hd}).
Unlike the up sector, where matter always couples to the same $ SO(10) $ VEV, each diagram like Fig.~\ref{fig:hd} involving a different flavon will couple to a different linear combination of VEVs.
This introduces CG factors non-trivially into $ Y^d $ and $ Y^e $.
As such, there is no fixed relationship between down-type quark and charged lepton Yukawa couplings, neither within a family, nor across families.
They are nevertheless expected to be of the same order.

One of the characteristic features of GUTs is the prediction of proton decay. 
It has not been observed and the proton lifetime is constrained to be $\tau_p > 10^{34} $ years \cite{Olive:2016xmw}. 
Proton decay can be mediated by the extra gauge bosons and by the triplets accompanying the Higgs doublets. 
In SUSY $SO(10)$ GUTs the main source for proton decay comes from the triplet Higgsinos. 
The decay width is dependent on SUSY breaking and the specific coupling texture of the triplets. 
In general the constraints are barely met when the triplets are at the GUT scale \cite{Nath:2006ut} (as shown in Appendix~\ref{sec:app:DTsplitting}, this is our case). 

The existence of additional fields in the model may allow proton decay from effective terms of the type
\begin{equation}
	gQQQL\frac{\braket{X}}{M_P^2}.
\end{equation}
Such terms must obey the constraint $ g \braket{X} < 3\times 10^{9} $ GeV.
In our model, the largest contribution of this type comes from the term
\begin{equation}
	\psi\psi\psi\psi
	\frac{H^{B-L}_{45}(H^{X,Y}_{45}H^Z_{45})^2}{M^6_P}
	\quad \Rightarrow \quad 
	\braket{X} = \frac{(M_\mathrm{GUT})^5}{M^4_P}\sim 10^{3} \mathrm{~GeV}.
\end{equation}
The constraint on $ \braket{X} $ is easily met, so proton decay from such terms is highly suppressed.

\section{Mass matrices, estimates and fits}
\label{sec:massmatrices}

\subsection{Mass matrices}
We present here the Yukawa and mass matrices, which will be used in the numerical fits below.
The full derivation of these matrices, taking into account $ S_4 $ products, CG factors and third-family mixing, is given in Appendix~\ref{sec:app:fullderivation}. 
We begin by defining numerical matrices
\begin{alignat}{3}
	Y_{11} &= \pmatr{1&1&3\\1&1&3\\3&3&9} , \quad &
	Y_{22} &= \pmatr{0&0&0\\0&1&1\\0&1&1} , \quad &
	Y_{33} &= \pmatr{0&0&0\\0&0&0\\0&0&1} , \nonumber \\[1ex]
	Y_{12} &= \pmatr{0&1&1\\1&2&4\\1&4&6} , \quad &
	Y_{P}  &= \pmatr{0&0&-1\\0&2&0\\-1&0&0} .&&
\label{eq:Ymatrices}
\end{alignat}
We note that all matrices derive from triplet products like $ (\psi \phi_i)(\psi \phi_j) $,
with $S_4$ singlet contractions in each bracket, except $ Y_P $ which derives from the Planck-suppressed operator $ \psi \psi \phi_3 H^d_{10} $.

The up, down, charged lepton and Dirac neutrino Yukawa matrices ($ Y^u $, $ Y^d $, $ Y^e $ and $ Y^\nu $ respectively) and RH neutrino mass matrix $ M^\mathrm{R} $ arising from Figs.~\ref{fig:hu}-\ref{fig:h16},
assuming that the MSSM Higgs doublets $h_u$ and $h_d$ arise from $H^u_{10}$ and $H^d_{10}$, respectively, as shown in Appendix~\ref{sec:app:DTsplitting}, may then be expressed, as shown in Appendix~\ref{sec:app:fullderivation}, as
\begin{alignat}{8}
	& Y^u 
	&&=\, & y^u_1 e^{i\eta} Y_{11} & 
	\, +\, & y^u_2 Y_{22} & 
	\, +\, & y^u_3 e^{i \eta^\prime} Y_{33} & 
	, \label{eq:Yu} \\ 
	& Y^\nu 
	&&=\, & y^\nu_1 e^{i\eta} Y_{11} &
	\, +\, & y^\nu_2 Y_{22} &
	\, +\, & y^\nu_3 e^{i \eta^\prime} Y_{33} & 
	, \label{eq:Ynu} \\
	&M^\mathrm{R}
	&&=\, & M^\mathrm{R}_1 e^{i\eta} Y_{11} & 
	\, +\, & M^\mathrm{R}_2 Y_{22} &  
	\, +\, & M^\mathrm{R}_3 e^{i \eta^\prime} Y_{33} & 
	, \label{eq:MR}\\
	& Y^d 
	&&=\, & y^d_{12} e^{i\frac{\eta}{2}} Y_{12} &
	\, +\, & y^d_2 e^{i\alpha_d} Y_{22} & 
	\, +\, & y^d_3 e^{i\beta_d} Y_{33} & 
	\, +\, & y^P e^{i\gamma} Y_{P}  &
	, \label{eq:Yd} \\ 
	& Y^e 
	&&=\, & y^e_{12} e^{i\frac{\eta}{2}} Y_{12} &
	\, +\, & y^e_2 e^{i\alpha_e} Y_{22} &
	\, +\, & y^e_3 e^{i \beta_e} Y_{33} &
	\, +\, & y^P e^{i\gamma} Y_{P} & 
	. \label{eq:Ye}
\end{alignat}
The flavon VEVs $ v_a $ are complex, with the fixed phase relation 
\begin{equation}
	\eta = \arg \left(\frac{v_1}{v_2}\right)^2 
	= - \frac{2\pi}{3},
\end{equation}
given (up to a sign) by the superpotential that fixes the alignments. 
The remaining phase $\eta^\prime$ is determined by the fit.

The light neutrino mass matrix is obtained by the seesaw mechanism. 
Both $ Y^\nu $ and $ M^\mathrm{R} $ have the same structure, namely both are sums over the same rank-1 matrices $ Y_{11} $, $ Y_{22} $ and $ Y_{33} $. 
By a proof given in \cite{Bjorkeroth:2016lzs}, the light neutrino matrix $ m^\nu $ will also have this structure, i.e.
\begin{equation}
\begin{split}
	m^\nu 
	&= \mu_1 e^{i\eta} Y_{11}
		+ \mu_2 Y_{22}
		+ \mu_3 e^{i \eta^\prime} Y_{33} \\
	&=
		\mu_1 e^{i\eta} \pmatr{1&1&3\\1&1&3\\3&3&9} 
		+ \mu_2 \pmatr{0&0&0\\0&1&1\\0&1&1} 
		+ \mu_3 e^{i \eta^\prime} \pmatr{0&0&0\\0&0&0\\0&0&1},
\end{split}
\end{equation}
where the parameters $ \mu_i $ are given in terms of the parameters $ y^\nu_i $ and $ M^\mathrm{R}_i $ simply by 
\begin{equation}
	\mu_i = v_u^2 \frac{(y^\nu_i)^2}{M^\mathrm{R}_i}.
\label{eq:mui}
\end{equation}
As shown in the introduction, the flavons yield a light neutrino mass matrix $ m^\nu $, where the normal hierarchy $ m_1\ll m_2 \ll m_3 $ then corresponds to $\mu_3 \simlt \mu_1 \ll \mu_2$. Achieving this hierarchy after seesaw implies that the RH neutrino masses are very hierarchical, as we will see below.%
\footnote{
While the model does not mathematically forbid an inverted hierarchy, we have checked that the corresponding predictions for neutrino masses and mixing angles would always give a bad fit to data. 
It would also require parameter choices that strongly violate the naturalness principle employed here.
}

\subsection{Analytic estimates}
\label{sec:analytics}

The mass matrices involve the following real free parameters: $ y^u_i $, $ y^d_i $, $ y^e_i $, $ \mu_i $, and $ y^P $ (a total of 13). 
Recalling that $ \eta $ is fixed by flavon vacuum alignment, we have the following further free parameters: $ \eta^\prime $, $ \alpha_{d,e} $, $ \beta_{d,e} $, and $ \gamma $ (a total of 6).
The scales of the real parameters are mostly fixed by the scales of the flavon VEVs, $ v_{1,2,3} $. 
We set the flavon VEV scales to some appropriate values,
\begin{equation}
	v_1 \approx 0.002 M_\mathrm{GUT} , \qquad
	v_2 \approx 0.05 M_\mathrm{GUT} , \qquad
	v_3 \approx 0.5 M_\mathrm{GUT} ,
\end{equation}
where we set $ M_\mathrm{GUT} \simeq 10^{16} $ GeV.
The terms giving $ M_{1,2}^\mathrm{R} $ and $ y^P $ in Eqs.~\ref{eq:MR} and \ref{eq:Yd}-\ref{eq:Ye}, respectively, derive from terms suppressed by one Planck mass $ M_P $.
As they arise from unspecified dynamics, the scale of these parameters is not very well defined. 
For definiteness, we set $ M_P \simeq 10^{19} $ GeV and again assume the associated coefficients are close to one. 
We assume that $M_{\rho} \sim M_\mathrm{GUT}$.

We may estimate the parameters of the matrices defined in Eqs.~\ref{eq:Yu}-\ref{eq:Ye} as follows: set all $ \mathcal{O}(1) $ coefficients to exactly one, and ignore CG factors by setting all adjoint Higgs VEVs to $ M_\mathrm{GUT} \simeq 10^{16} $ GeV. 
Then the Yukawa couplings are estimated to be
\begin{equation}
\arraycolsep=1pt
\def\arraystretch{1.3}
\begin{array}{rcccccccccl}
	&& &&
	y^u_1 &\sim& y^\nu_1 &\sim& {v_1^2}/{M_\mathrm{GUT}^2} &\approx& 4 \times 10^{-6} , \\
	y^u_2 &\sim& y^\nu_2 &\sim& y^d_2 &\sim& y^e_2 &\sim& {v_2^2}/{M_\mathrm{GUT}^2} &\approx& 2.5 \times 10^{-3} , \\
	y^u_3 &\sim& y^\nu_3 &\sim& y^d_3 &\sim& y^e_3 &\sim& {v_3^2}/{M_\mathrm{GUT}^2} &\approx& 0.25, \\
	&& && y^d_{12} &\sim& y^e_{12} &\sim& v_1 v_2 /M_\mathrm{GUT}^2 &\approx& 1 \times 10^{-4} , \\
	&& && && y^P &\sim& v_3 / M_P &\approx& 5 \times 10^{-4}.
\end{array}
\end{equation}
The RH neutrino mass parameters are estimated to be
\begin{equation}
	M^\mathrm{R}_1 \sim 4 \times 10^{7} \mathrm{~GeV} , \quad
	M^\mathrm{R}_2 \sim 2.5 \times 10^{10} \mathrm{~GeV} , \quad   
	M^\mathrm{R}_3 \sim 10^{16} \mathrm{~GeV} .
\end{equation}
This very strong hierarchy implies negligible RH neutrino mixing, such that the mass eigenvalues closely correspond to the above values.
As each parameter contains several $ \mathcal{O}(1) $ coefficients $ \lambda $ and CG factors, the above numbers only represent order of magnitude estimates.

As we will see in the numerical fit below, the above estimates are in good agreement with the values that produce a good fit to data, with a single exception: the parameter $ M^\mathrm{R}_1 $, which is primarily responsible for the lightest RH neutrino mass, should be a factor $\mathcal{O}(0.01)$ times the estimate above in order to give the correct light neutrino mass spectrum. 
This can be understood by inserting the above estimates for $ y^\nu_1 $ and $ M^\mathrm{R}_1 $ into the expression for $ \mu_1 $ in Eq.~\ref{eq:mui}, which suggests $ \mu_1 \sim 0.01 $ meV, whereas we will see the fit prefers a value of $ \mathcal{O}(1) $ meV.
The necessary factor can be achieved by assuming one or more coefficients deviates from unity.

One may also obtain approximate expressions for the quark mixing angles in terms of quark Yukawa couplings as follows. 
The very strong hierarchy in the three real parameters of $ Y^u $ is correlated with that in the physical Yukawa eigenvalues of up, charm and top quarks. 
We therefore expect negligible contributions from the up sector to quark mixing. 
This implies that not only do the four real parameters in the down sector, $ y^d_i $ and $ y^P $, fix the down-type Yukawa eigenvalues, they also must reproduce the observed CKM mixing angles. 

Let us consider $ Y^d $, keeping only the leading terms in each element. For simplicity, we ignore free phases.
As noted above, $ y^d_{12} \sim y^P < y^d_2 \ll y^d_3 $. 
We also define $ y_2^\prime = y^d_2 + 2 y^d_{12} + 2y^P $. Then
\begin{equation}
	Y^d \approx 
	\pmatr{
		0 & y^d_{12} & y^d_{12}-y^P \\
		y^d_{12} & y_2^\prime & y_2^\prime + 2(y^d_{12}-y^P) \\
		y^d_{12}-y^P & y_2^\prime + 2(y^d_{12}-y^P) & y^d_3
	} .
\end{equation}
In the small angle approximation, the mixing angles can be estimated by
\begin{equation}
\begin{split}
	\theta_{12}^q \approx \frac{Y^d_{12}}{Y^d_{22}} 
	= \frac{y^d_{12}}{y_2^\prime} , \quad
	\theta_{13}^q \approx \frac{Y^d_{13}}{Y^d_{33}} 
	= \frac{y^d_{12}-y^P}{y^d_3} , \quad
	\theta_{23}^q \approx \frac{Y^d_{23}}{Y^d_{33}} 
	= \frac{y_2^\prime+2(y^d_{12}-y^P)}{y^d_3} .
\end{split}
\end{equation}
The down-type Yukawa eigenvalues are given by
$ y_d \approx (y^d_{12})^2/y_2^\prime $,
$ y_s \approx y_2^\prime $,
$ y_b \approx y^d_3 $.
Solving for $ y^d_{12}$, $ y_2^\prime $ and $y^d_3 $, we have, to good approximation,
$ y^d_{12} \approx \sqrt{y_d y_s} $,
$ y_2^\prime \approx y_s $,
$ y^d_3 \approx y_b $.
Reintroducing these into our estimates for mixing angles, we get
\begin{equation}
	\theta_{12}^q \approx \sqrt{\frac{y_d}{y_s}}, \quad
	\theta_{13}^q \approx \frac{\sqrt{y_d y_s}-y^P}{y_b} , \quad
	\theta_{23}^q \approx \frac{y_s + 2 (\sqrt{y_s y_d}-y^P)}{y_b} .
\label{eq:quarkanglesapprox}
\end{equation}

Note that the first equality is exactly the GST relation \cite{Gatto:1968ss}, which is in good agreement with data.
In fact, the GST relation, which predicts $ \theta^q_{12} \simeq 0.224 $ for the central values of $ y_d $ and $ y_s $, is in mild tension with experimental data, which gives $ \theta^q_{12} \simeq 0.227 $. 
Possible modifications to the GST result have been proposed \cite{Ross:2009zz}, e.g. adding a correction like $ \sqrt{y_u/y_c} $, which can be realised by a texture zero also in $ Y^u $. 
Alternatively, one may exploit the statistical uncertainties on each of the down and strange quark masses. 
A small deviation from their central values can predict a slightly different $ \theta_{12}^q $.

On the other hand, the mixing angles $ \theta_{13}^q $ and $ \theta_{23}^q $ are less precisely estimated, as the parameter $ y^P $ can be as large as $ y^d_{12} $, and the final result will depend on the relative phase between $ y^d_{12} $ and $ y^P $.  
Note however that both mixing angles depend in the same way on $ y^d_{12} - y^P $. 
Generally, the approximations in Eq.~\ref{eq:quarkanglesapprox} predict some tension between $ \theta_{13}^q $ and $ \theta_{23}^q $, which are too large and too small, respectively. 
This tension cannot be resolved simply by tuning $ y^P $.

\subsection{Numerical fit}
\label{sec:numerics}

Our model determines the Yukawa couplings and mixing parameters at the GUT scale, which is also the highest flavour-breaking scale. 
The values from experiments must therefore be run up to the GUT scale. 
Moreover, when matching the SM to the MSSM at the scale $ M_{\mathrm{SUSY}} $, supersymmetric radiative threshold corrections have to be included. 
Such an analysis has been performed in \cite{Antusch:2013jca}.
The parametrisation of these corrections is summarised in Appendix~\ref{sec:app:susy}. 
Most parameters do not significantly affect the fit, so are simply set to reasonable values. 
Specifically, we set $ M_\mathrm{SUSY} = 1 $ TeV, $ \tan \beta = 5 $ and $ \bar{\eta}_q = \bar{\eta}_\ell = 0 $. 
We also find that a good fit can be achieved for a rather large value $ \bar{\eta}_b = -0.8 $. 
The choices of SUSY parameters $ \tan \beta $ and $ \bar{\eta}_b $ are here empirically determined to give a good fit of the model to data.
It is clear from the fit that large (negative) $ \bar{\eta}_b $ is required, affecting primarily the bottom quark Yukawa coupling $ y_b $.
In order to keep $ y_b $ perturbative, we must assume reasonably small $ \tan \beta $. 
In the region of $ 5 < \tan \beta < 10 $ or so, the fit is rather insensitive to the exact choice.
Neutrino data is taken from the NuFit global fit \cite{Esteban:2016qun}.

To find the best fit of the model to data, we minimise a $ \chi^2 $ function, defined in the standard way: for a given set of input parameters $ x $, we calculate the $ n $ observables $ P_n(x) $. 
These are then compared to the observed values $ P^\mathrm{obs}_n $, which have associated statistical errors $ \sigma_n $.%
\footnote{
	In order for a minimum $ \chi^2 $ to correspond to the maximum likelihood, the statistical uncertainties should be symmetric (Gaussian). This is essentially satisfied for all parameters except $ \theta_{23}^\ell $, where current experimental data cannot conclusively resolve the octant, i.e whether it is larger or smaller than $ 45^\circ $. Currently, the data favours $ \theta_{23}^\ell < 45^\circ $, with a central value $41.6^\circ$ \cite{Esteban:2016qun}. We will assume this is the true value.
}
Then 
\begin{equation}
	\chi^2 = \sum_n \left( \frac{P_n(x) - P^\mathrm{obs}_n}{\sigma_n} \right)^2.
\end{equation}
For our model, the input parameters are
$ x = \{ y^u_i, y^d_i, y^e_i, y_P, \mu_i, \eta^\prime, \alpha_{d,e}, \beta_{d,e}, \gamma \} $,
and the observables are given by 
$ P_n \in \{ \theta^q_{ij}, \delta^q, y_{u,c,t}, y_{d,s,b}, \theta^\ell_{ij}, y_{e,\mu,\tau}, \Delta m_{ij}^2 \} $.
Note that as the lepton $CP$ phase $ \delta^\ell $ is not yet well measured, we do not include it in the fit, rather we prefer to leave it as a pure prediction. Furthermore, only the neutrino mass-squared differences are measured in oscillation experiments (as opposed to the masses themselves), while our model predicts the masses outright, including the lightest neutrino mass $ m_1 $.

\begin{table}[!ht]
\centering
\footnotesize
\renewcommand{\arraystretch}{1.1}
\begin{tabular}{ l cc c cc }
\toprule
	\multirow{2}{*}{Observable}& \multicolumn{2}{c}{Data} && \multicolumn{2}{c}{Model} \\
\cmidrule{2-3} \cmidrule{5-6}
	& Central value & 1$\sigma$ range  && Best fit & Interval \\
\midrule
	$\theta_{12}^\ell$ $/^\circ$ & 33.57 & 32.81 $\to$ 34.32 && 33.62 & 31.69 $\to$ 34.46 \\ 
	$\theta_{13}^\ell$ $/^\circ$ & 8.460 & 8.310 $\to$ 8.610 && 8.455 & 8.167 $\to$ 8.804 \\  
	$\theta_{23}^\ell$ $/^\circ$ & 41.75 & 40.40 $\to$ 43.10  && 41.96 & 39.47 $\to$ 43.15 \\ 
	$\delta^\ell$ $/^\circ$ & 261.0 & 202.0 $\to$ 312.0 && 300.9 & 280.7 $\to$ 308.4 \\
	$y_e$  $/ 10^{-5}$ & 1.017 &  1.011 $\to$ 1.023 && 1.017 & 1.005 $\to$ 1.029 \\ 
	$y_\mu$  $/ 10^{-3}$ & 2.147 & 2.134 $\to$ 2.160 && 2.147 & 2.121 $\to$ 2.173 \\ 
	$y_\tau$  $/ 10^{-2}$ & 3.654 & 3.635 $\to$ 3.673 && 3.654& 3.616 $\to$ 3.692 \\ 
	$\Delta m_{21}^2 / (10^{-5} \, \mathrm{eV}^2 ) $ & 7.510  & 7.330 $\to$ 7.690 && 7.515 & 7.108 $\to$ 7.864 \\
	$\Delta m_{31}^2 / (10^{-3} \, \mathrm{eV}^2) $ & 2.524  & 2.484 $\to$ 2.564 && 2.523 & 2.443 $\to$ 2.605 \\
	$m_1$ /meV & & && 0.441 & 0.260 $\to$ 0.550 \\ 
	$m_2$ /meV & & && 8.680 & 8.435 $\to$ 8.888 \\ 
	$m_3$ /meV & & && 50.24 & 49.44 $\to$ 51.05 \\
	$\sum m_i$ /meV & \multicolumn{2}{c}{$<$ 230} && 59.36 & 58.49 $\to$ 60.19 \\
	$ \alpha_{21} $ & & && 67.90 & $-25.19$ $\to$ 87.49\phantom{+} \\
	$ \alpha_{31} $ & & && 164.2 & 19.98 $\to$ 184.5 \\
\bottomrule     		
\end{tabular}
\caption{Model predictions in the lepton sector for $\tan \beta = 5$, $ M_{\mathrm{SUSY}} = 1 $ TeV and $ \bar{\eta}_b = -0.8 $. The observables are at the GUT scale. The lepton contribution to the total $\chi^2$ is 0.03. $\delta^\ell$ as well as the neutrino masses $m_i$ are pure predictions of our model. The model interval is a Bayesian 95\% credible interval. The bound on $ \sum m_i $ is taken from \cite{Ade:2015xua}.}
\label{tab:numleptons}
\end{table}

\begin{table}[!ht]
\centering
\footnotesize
\renewcommand{\arraystretch}{1.1}
\begin{tabular}{ l cc c cc }
\toprule
	\multirow{2}{*}{Observable} & \multicolumn{2}{c}{Data} && \multicolumn{2}{c}{Model} \\
\cmidrule{2-3} \cmidrule{5-6}
	& Central value & $1\sigma$ range && Best fit & Interval \\
\midrule
	$\theta_{12}^q$ $/^\circ$ &13.03 & 12.99 $\to$ 13.07 && 13.02 & 12.94 $\to$ 13.10 \\	
	$\theta_{13}^q$ $/^\circ$ &0.039 & 0.037 $\to$ 0.040 && 0.039 & 0.036 $\to$ 0.041 \\
	$\theta_{23}^q$ $/^\circ$ &0.445& 0.438 $\to$ 0.452 && 0.439 & 0.426 $\to$ 0.450 \\	
	$\delta^q$ $/^\circ$ & 69.22 & 66.12 $\to$ 72.31  && 69.21 & 63.22 $\to$ 73.94 \\
	$y_u$  $/ 10^{-6}$ & 2.988 & 2.062 $\to$ 3.915 && 3.012 & 1.039 $\to$ 4.771 \\	
	$y_c$  $/ 10^{-3}$ & 1.462 & 1.411 $\to$ 1.512 && 1.493 & 1.445 $\to$ 1.596 \\	
	$y_t$  			   & 0.549 & 0.542 $\to$ 0.556 && 0.547 & 0.532 $\to$ 0.562 \\
	$y_d$  $/ 10^{-5}$ & 2.485 & 2.212 $\to$ 2.758 && 2.710 & 2.501 $\to$ 2.937 \\	
	$y_s$  $/ 10^{-4}$ & 4.922 & 4.656 $\to$ 5.188 && 5.168 & 4.760 $\to$ 5.472 \\
	$y_b$   		   & 0.141 & 0.136 $\to$ 0.146 && 0.137 & 0.126 $\to$ 0.143 \\
\bottomrule
\end{tabular}
\caption{Model predictions in the quark sector for  $\tan\beta = 5$, $M_{\mathrm{SUSY}} = 1$ TeV and $\bar{\eta}_b = -0.8$. The observables are at the GUT scale. The quark contribution to the total $\chi^2$ is 3.38. The model interval is a Bayesian 95\% credible interval.}
\label{tab:numquarks}
\end{table}

\begin{figure}[ht]
	\centering
	\includegraphics[width=0.8\textwidth]{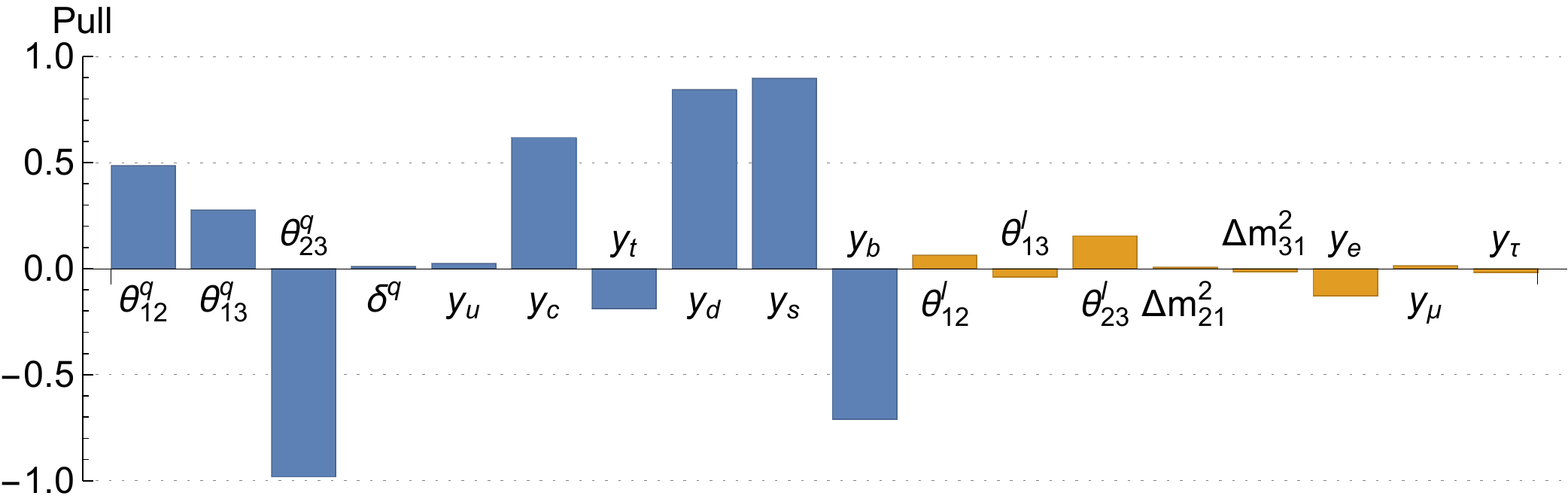}
	\caption{Pulls for the best fit of model to data, as shown in Tables \ref{tab:numleptons}-\ref{tab:numquarks}, for quark (blue) and lepton (yellow) parameters.}
\label{fig:pulls}
\end{figure}

\begin{table}[ht]
\centering
\footnotesize
\renewcommand{\arraystretch}{1.1}
\begin{tabular}[t]{lr}
\toprule
	Parameter & Value \\ 
\midrule
	$y^u_1 \, /10^{-6}$ & 3.009 \\
	$y^u_2 \, /10^{-3}$ & 1.491 \\
	$y^u_3$ & 0.549 \\
	$y^d_{12} \, /10^{-4}$ &$-1.186$ \\
	$y^d_2 \, /10^{-4}$ & 6.980 \\
	$y^d_3$ & 0.137 \\
	$y^P \, /10^{-4}$ & 1.243 \\ 
\bottomrule
\end{tabular}
\hspace*{0.5cm}
\begin{tabular}[t]{lr}
\toprule
	Parameter & Value \\ 
\midrule
	$y^e_{12}$ $/10^{-4}$ & 1.558 \\
	$y^e_2$ $/10^{-3}$ & 2.248 \\
	$y^e_3$ $/10^{-2}$ & 3.318 \\
	$\mu_1 $ /meV & 2.413 \\
	$\mu_2 $ /meV & 27.50 \\
	$\mu_3 $ /meV & 2.900 \\
\bottomrule
\end{tabular}
\hspace*{0.5cm}
\begin{tabular}[t]{lr}
\toprule
	Parameter & Value \\ 
\midrule
	$\alpha_d$ & $ 0.043\pi $  \\
	$\beta_d$ & $ 0.295\pi $ \\ 
	$\alpha_e$ & $ 1.692\pi $ \\ 
	$\beta_e$ &  $ 1.755\pi $ \\
	$\gamma$& $ 0.918\pi $ \\ 
	$\eta^\prime$ & $ 1.053\pi $ \\
\bottomrule
\end{tabular}
\caption{Best fit input parameter values. The model has 13 real parameters: $y^u_i$, $y^d_i$, $y^e_i$, $\mu_i$ and $y^P$. While $\eta$ is fixed by flavon alignment to $-2\pi/3$, there are six additional free phases: $\eta^\prime$, $\alpha_{d,e}$, $\beta_{d,e}$ and $\gamma$. The total $\chi^2$ is 3.4.} 
\label{tab:parameters}
\end{table}

We present the best fit (minimum $ \chi^2 $) of the model to physical observables (Yukawa couplings and neutrino mass and mixing parameters) in Tables \ref{tab:numleptons} and \ref{tab:numquarks}, which also include the central values and $ 1\sigma $ ranges from data.
Fig.~\ref{fig:pulls} shows the associated pulls, and Table~\ref{tab:parameters} shows the corresponding input parameter values. 
The fit gives $\chi^2 \approx 3.4$.%
\footnote{
	The best fit predicts a strong neutrino hierarchy, with $ m_1 < 1 $ meV. 
	It is possible to achieve a milder hierarchy, although the numerical fit gives $ \chi^2 \simgt 20 $ in such cases, predicting neutrino masses of approximately $ 5 $, $ 10 $ and $ 51 $ meV. 
	Additionally it predicts $ \delta^\ell \approx +25^\circ $, currently disfavoured by experiment. 
}
A second minimum with $ \chi^2 \approx 4 $ was also found, leading primarily to a different prediction for $ \delta^\ell $, as discussed below, although we shall not present the full fit parameters for this case.

We see from Tables~\ref{tab:numleptons}, \ref{tab:numquarks} and Fig.~\ref{fig:pulls} that both quark and lepton sectors are fitted to within $1\sigma$ of the values predicted by global fits to experiment. 
The biggest pulls are in down-type quark Yukawa couplings $ y_{d,s,b} $ and $\theta^q_{23}$.
As shown in Section \ref{sec:analytics}, $\theta^q_{23}$ is approximately given by the ratio $ y_s / y_b $, which is typically too small. 
Furthermore, attempts to increase $ \theta_{23}^q $, e.g. by tuning $ y^P $, tends to increase $ \theta_{13}^q $, which is then too large.
This tension can be ameliorated by assuming large threshold corrections, i.e. by setting $ \bar{\eta}_b = -0.8 $, although some tension remains among the above parameters, which deviate by about $ 1\sigma $.

Tables~\ref{tab:numleptons} and \ref{tab:numquarks} also include a Bayesian 95\% credible interval for each observable,%
\footnote{
	This is analogous to, but should not be confused with, a frequentist confidence interval.
}
which was found by performing a Markov Chain Monte Carlo (MCMC) analysis. 
The interval for a given parameter corresponds to the region of highest posterior (probability) density (hpd), marginalised over the other parameters, and may be interpreted as follows: given the data, there is a 95\% probability that the true model value of that observable resides in the stated interval. 
For many observables, these probability distributions are essentially Gaussian, centred around the best fit value.
This is not always the case: the distributions for $ \theta^\ell_{12} $ and $ \theta^\ell_{23} $ are asymmetric, consisting of two partially overlapping peaks.
Moreover, the hpd region for $ \delta^\ell $ consists of two completely distinct intervals, which contain the best fit values $ 300.9^\circ $ (as seen in Table \ref{tab:numleptons}) and $233.9^\circ$ (corresponding to a second best fit point with $ \chi^2 \approx 4 $). 
Their associated 95\% credible intervals are given by 
$ 280.7 < \delta^\ell < 308.3 $ and
$ 225.1 < \delta^\ell < 253.2 $,
respectively.
We note that neither region includes maximal $CP$ violation $ \delta^\ell = 270^\circ $, which is close to the prediction from CSD3 with diagonal charged leptons. 
In short, charged-lepton corrections induce a deviation from maximal $CP$ phase, which can either be positive or negative, depending on the phases of $ Y^e $.

One may be tempted to calculate a reduced chi-squared $ \chi^2_\mathrm{red} $, i.e. the $ \chi^2 $ per degree of freedom (d.o.f.), where the number of d.o.f. is naively given by the number of observables minus the number of input parameters. 
In the conventional picture, a good fit has $ \chi^2_\mathrm{red} \simeq 1 $.
However, as discussed in \cite{Andrae:2010gh}, this interpretation is only valid for linear models, which our model is not. 
Indeed, when evaluating $ \chi^2 $ we fit 19 inputs to 18 observables, which in a linear model would suggest a perfect fit is always possible; this is certainly not the case. 
While $ \chi^2 $ is a valid tool for comparing models to each other, since it is not possible to establish an exact number of d.o.f., we cannot reliably define $ \chi^2_\mathrm{red} $.

\section{Conclusion}
\label{sec:conclusion}

The flavour puzzle in the SM is the source of a majority of the SM free parameters, characterised by different mixing behaviours for quarks and leptons, and very hierarchical masses.
The most minimal solution to the problem of neutrino masses remains the seesaw mechanism with heavy RH neutrinos, which arise automatically in $ SO(10) $, with naturally large masses. 
This motivates $ SO(10) $ above other popular gauge groups, such as $ SU(5) $, where RH neutrinos are added by hand. 
All three families of SM fermions in the \textbf{16} of $ SO(10) $ are here also unified in a single triplet of $ S_4 $. 
This very elegant picture presents model-building challenges, many of which we have tackled in this paper. 

We have constructed a rather simple, natural and complete $ SO(10) $ model of flavour with a discrete $ S_4 \times \mathbb{Z}_4^2 \times \mathbb{Z}_4^R $ symmetry, where all Yukawa matrices derive from the VEVs of triplet flavons, in the CSD3 alignment. 
It is simple in the sense that the field content is reasonably minimal, with small Higgs representations of $SO(10)$ consisting of two \textbf{10}s which contain the MSSM doublets, a Higgs spinor pair \textbf{16} and $\overbar{\mathbf{16}}$ responsible for Majorana masses and four adjoint Higgs \textbf{45}s, which provide necessary Clebsch-Gordan factors that distinguish charged leptons and down-type quarks. 
It is natural in the sense that Yukawa and mass matrices consist of sums of low-rank matrices, each of which contributes dominantly to a particular family, i.e. ``universal sequential dominance''. 
It is complete in the sense that we address the $\mu$-problem, Higgs mixing and doublet-triplet splitting, and provide an ultraviolet renormalisable model, with Planck-suppressed operators controlled by symmetry. 
However, we do not discuss the origin of the hierarchy of flavon VEVs, nor do we repeat the discussion of flavon vacuum alignment, which can be found in \cite{King:2016yvg}.

We believe this model represents a signficant step forward in the quest for a complete and correct description of fermions within SUSY GUTs. 
For instance, we have demonstrated the correct procedure for treating the third family couplings and how to generate an electroweak-scale renormalisable third-family Yukawa coupling.
We also emphasise that the principle of universal sequential dominance is a simple and effective way to understand fermion hierarchies.
Although the origin of such family hierarchies has not been fully resolved, as the scales of flavon VEVs $ v_i $ are assumed rather than proven,
the problem has been ameliorated, since the hierarchy is given by the squares of these VEVs. 

The model successfully reproduces the observed fermion masses and mixing, even in the quark sector, where the CKM parameters are measured to very high precision. 
Analytical estimates are underpinned by a detailed numerical analysis, demonstrating the viability of the model.
Moreover, there is no tuning of $ \mathcal{O}(1) $ parameters necessary to explain the mass hierarchies of charged fermions, accounting also for the milder hierarchy in down-type quarks compared to up-type quarks. 
The model simultaneously realises large lepton mixing and small quark mixing, as well as the GST relation for the Cabibbo angle, $ \theta^q_{12} \approx \sqrt{y_d/y_s} $ via a texture zero in the down-type Yukawa matrix $ Y^d $.
In the lepton sector an excellent fit to data is found, predicting a normal neutrino hierarchy and lightest neutrino mass $ m_1 \simlt 0.5 $ meV. 
The $CP$ phase $ \delta^\ell $ was not fitted, but left as a pure prediction. 
Two distinct regions are preferred, with corresponding best fit values $ \delta^\ell \approx 301^\circ $ and $ 234^\circ $. 
We emphasise that the model predicts significant deviation from both zero and maximal $CP$ violation.

\subsection*{Acknowledgements}

S.\,F.\,K. acknowledges the STFC Consolidated Grant ST/L000296/1.
This project has received funding from the European Union's Horizon 2020 research and innovation programme under the Marie Sk\l{}odowska-Curie grant agreements 
Elusives ITN No.\ 674896 and
InvisiblesPlus RISE No.\ 690575.
F.\,B. thanks Mathew Smith and Robert Firth for helpful discussions.

\appendix
\section{Doublet-triplet and doublet-doublet splitting}
\label{sec:app:DTsplitting}

As is the case for every broken GUT, the Higgs sector of our model contains more fields than the usual MSSM. 
The $H_{10}^{u,d}$ multiplets contain colour triplets that mediate proton decay. Since we have two \textbf{10}s, there is an additional pair of doublets that, if light, could spoil gauge coupling unification. 
For these reasons, those extra fields need to be heavy, while ensuring the MSSM doublets are massless. 
This splitting can be achieved in our model.

The splitting mechanism involves superfields given in Table~\ref{tab:funfields}. 
The singlet $\xi$ obtains a VEV slightly above the GUT scale and ensures the correct structure to the masses. 
The $H_{16}$ generates a mass for the $H_{\overbar{16}}$ and also gets a VEV in the RH neutrino ($ \nu^c $) direction. 
$H_{45}^{B-L}$ is the only $R$-charged field that gets a VEV, breaking $\mathbb{Z}_4^R$ to the usual $R$ parity. 
This splitting mechanism needs three extra messenger pairs 
which are listed in table \ref{ta:Hmen}.

\begin{table}[ht]
\centering
	\begin{tabular}{ l c@{\hskip 5pt} c c c c}
	\toprule
	\multirow{2}{*}{\rule{0pt}{4ex}Field}	& \multicolumn{5}{c}{Representation} \\
	\cline{2-6}
	\rule{0pt}{3ex}
	& $S_4$ & $SO(10)$ &$\mathbb{Z}_{4}$ & $\mathbb{Z}_{4}$ &  $\mathbb{Z}_{4}^R$ \\
	\midrule
	$\xi$ & 1 & 1 & 2 & 2 & 0 \\
	\rule{0pt}{3ex}%
	$\bar{\chi}_u$ & 1 & $\overbar{16}$ & 2 & 1 & 2 \\
	${\chi}_u$ & 1 & 16 & 0 & 1 & 0 \\
	$\bar{\chi}_d$ & 1 & $\overbar{16}$ & 1 & 0 & 0 \\
	${\chi}_d$ & 1 & 16 &1 & 2 & 2 \\
	\rule{0pt}{3ex}%
	$\zeta_1$ & 1 & 45 & 1 & 1 & 2 \\
	$\zeta_2$ & 1 & 45 & 1 & 1 & 0 \\
	\bottomrule
	\end{tabular}
\caption{Messengers involved in doublet-triplet splitting.
}
\label{ta:Hmen}
\end{table}

With them, we may write the superpotential (ignoring dimensionless couplings)
\begin{equation}\begin{split}
\mathcal{W}_H &= 
	H^{B-L}_{45} \left(H^u_{10}H^d_{10}+\zeta_2 \zeta_2+H_{\overbar{16}}\chi_u+H_{16}\overbar{\chi}_d\right)\\
	&\quad +H_{\overbar{16}}H_{10}^u\overbar{\chi}_u+H_{16}H_{10}^d\chi_d+H_{16}H_{\overbar{16}}\zeta_1	
	+\xi\left(\zeta_1\zeta_2+\overbar{\chi}_u\chi_u+\overbar{\chi}_d\chi_d\right) \\ 
	&\quad 
	+ H^{B-L}_{45}
	\left( 
	\frac{H_{\overbar{16}}H_{\overbar{16}}H^d_{10}}{M_P}
	+\frac{H_{16}H_{16}H^u_{10}}{M_P}
	+H_{10}^uH_{10}^d\frac{(H_{45}^{X,Y,Z})^4}{M_P^4}
	\right),
\end{split}\end{equation}
where we assume that the VEV $\braket{\xi} \simgt M_\mathrm{GUT}$, so that we may integrate out the messenger fields and obtain effective superpotential
\begin{equation}
\begin{split}
	\mathcal{W}_H &= H^{B-L}_{45}
	\left(
	H^u_{10}H^d_{10}
	+\frac{(H_{16}H_{\overbar{16}})^2}{\braket{\xi}^2}
	+\frac{H_{\overbar{16}}H_{\overbar{16}}H^u_{10}}{\braket{\xi}}
	+\frac{H_{16}H_{16}H^d_{10}}{\braket{\xi}}
	\right. \\ &\qquad 
	+ \left. 
	\frac{H_{\overbar{16}}H_{\overbar{16}}H^d_{10}}{M_P}
	+\frac{H_{16}H_{16}H^u_{10}}{M_P}
	+H_{10}^uH_{10}^d\frac{(H_{45}^{X,Y,Z})^4}{M_P^4}
	\right),
\label{eq:effhi}
\end{split}
\end{equation}
where we have suppressed dimensionless couplings,
and the final term involves all combinations of adjoints allowed by the symmetries, i.e. either $ (H^Z_{45})^4 $ or any combination of powers of $ H^X_{45} $ and $ H^Y_{45} $ totalling four.
The three terms suppressed by $\braket{\xi}$ are allowed by the integration of three messenger pairs. 

We assume that the superfields $H_{\overbar{16},16} $, $ H_{45}^k $ ($ k = X,Y,Z,B-L $) get GUT-scale VEVs, i.e. $ v_{16,\overbar{16}} \approx v_{45}^k \approx M_\mathrm{GUT}$, through an unspecified mechanism. 
$H_{\overbar{16},16}$ get VEVs in the $ \nu^c $ direction. 
$H_{45}^{B-L}$ gets a VEV aligned in the $B-L$ direction, which splits doublet and triplet Higgs masses through the Dimopoulos-Wilczek (DW) mechanism \cite{DW}. 
This can be understood by considering the decomposition of the $H_{10}^{u,d}$ into the Pati-Salam group. 
The triplets behave as a sextuplet of $SU(4)$ while the doublets are singlets. 
Since $U(1)_{B-L}\subset SU(4)$, the triplets get a mass from the first term of Eq.~\ref{eq:effhi} while the doublets don't. 
In the last term, all the $SO(10)$ adjoints can be contracted to a singlet, so they affect doublets and triplets equally.

To demonstrate the mechanism, we construct the doublet and triplet mass matrices. 
We define some dimensionless scale parameters $ y = M_\mathrm{GUT}/M_P$, $ z = M_\mathrm{GUT} / \braket{\xi} $.
We label the up-type doublets inside a given Higgs representation $H$ by $\textbf{2}_{u}(H) $, and down-type doublets by $\textbf{2}_{d}(H) $. 
We define triplets $\textbf{3}_{u}(H) $ and $\textbf{3}_{d}(H) $ analogously. 
$ H $ can be either $ H_{10}^u $, $ H_{10}^d $ or $ H_{\overbar{16},16} $.
The mass matrices $ M_D $ and $ M_T $ are given by
\begin{equation}
\begin{split}
	M_D &= 
	\begin{blockarray}{c ccc}
	& \textbf{2}_{u}(H^u_{10}) 
	& \textbf{2}_{u}(H^d_{10}) 
	& \textbf{2}_{u}(H_{\overbar{16}})\\[1ex]
	\begin{block}{c(ccc)}
	 \textbf{2}_{d}(H^d_{10})	& y^4	& 0 	& y  \\[0.5ex]
	 \textbf{2}_{d}(H^u_{10}) 	& 0 	& -y^4 	& z  \\[0.5ex]
	 \textbf{2}_{d}(H_{16}) 	& y		&  z 	& z^2\\
	\end{block}
	\end{blockarray} \ M_\mathrm{GUT} \,,
	\\
	M_T&= \begin{blockarray}{c ccc}
	& \textbf{3}_{u}(H^u_{10}) &  \textbf{3}_{u}(H^d_{10}) &  \textbf{3}_{u}(H_{\overbar{16}})\\[1ex]
	\begin{block}{c(ccc)}
	 \textbf{3}_{d}(H^d_{10}) & 1 &0&y \\[0.5ex]
	 \textbf{3}_{d}(H^u_{10}) &0 & -1 &  z \\[0.5ex]
	 \textbf{3}_{d}(H_{16}) & y  &  z &  z^2\\
	\end{block}
	\end{blockarray} \ M_\mathrm{GUT}\,.
\end{split}
\end{equation}
The triplets mass matrix $M_T$ has three eigenvalues of $ \mathcal{O}(M_\mathrm{GUT}) $.
The doublets mass matrix has two eigenvalues at $ \mathcal{O}(M_\mathrm{GUT}) $ and one at $ \mathcal{O}(y^4 M_\mathrm{GUT}) $, which we identify with the $\mu$ term. 
Since $ y \approx 10^{-3} $ we have $\mu\sim 1 $ TeV, which is the desired order. 
Furthermore, the light eigenvectors of $M_D$ define the MSSM doublets $h_{u,d}$ as
\begin{equation}
	h_u \approx \textbf{2}_{u}(H^u_{10}) 
	+ \frac{ y}{z}\textbf{2}_{u}(H^d_{10}),\quad
	h_d \approx \textbf{2}_{d}(H^d_{10}) 
	+ \frac{y}{z}\textbf{2}_{u}(H^d_{10}),
\end{equation}
where the contribution of $\mathcal{O}(y)$ is negligible, so that the MSSM doublets are located as required by the Yukawa structure of the model.

\section{Renormalisability of the third family}
\label{sec:app:third}

In this section we show that naive integration over messenger fields is not possible for the third family, due to the large VEV of $ \phi_3 $.
We reiterate that there is an assumed hierarchy of flavon VEVs, such that $ v_1 \ll v_2 \ll v_3 \sim M_\mathrm{GUT} $, implying it is not possible to formally integrate out the messengers $\chi_3$ which couple to the flavon $\phi_3$.

To explore this further, let us single out the terms in $ W_Y $ involving these fields and $ H^u_{10} $ (the same method applies to terms coupling to $ H^d_{10} $). 
Suppressing $ \mathcal{O}(1) $ couplings, the relevant terms are
\begin{equation}
	W_Y^{(3)} =
	\psi \phi_3 \overbar{\chi}_3 
	+ H_{45}^Z \chi_3 \overbar{\chi}_3 
	+ \chi_3 \chi_3 H_{10}^u.
\end{equation}
After fields acquire VEVs (with $ \braket{\phi_3} = v_3 (0,0,1) $), we have
\begin{equation}
	W_Y^{(3)} = 
	 v_3 \psi_3 \overbar{\chi}_3 + \braket{H_{45}^Z} \chi_3 \overbar{\chi}_3 .
\label{eq:WY3}
\end{equation}
These two terms are of comparable order.

Naively, $ \psi_3 $ may be interpreted as the set of third-family particles.
The problem with this picture is that it has a large coupling to $ \overbar{\chi}_3 $, which induces a mass for $ \psi_3 $ via the second term in Eq.~\ref{eq:WY3}. 
This clearly does not correspond to the physical third-family states (top quark and third Dirac neutrino), which are massless above the electroweak scale. 
To obtain the physical (massless) states, which we label $ t $, we rotate into a physical basis $ (\psi_3,\chi_3) \to (t,\chi) $, such that $ t $ does not couple to $ \overbar{\chi}_3 $. 
This basis change is given by
\begin{equation}
\begin{split}
	\psi_3 = 
	\frac{\braket{H_{45}^Z} t + v_3 \, \chi}{r}
	,
	\quad
	\chi_3 = 
	\frac{- v_3 \, t + \braket{H_{45}^Z} \chi}{r}
	;
	\quad
	r = \sqrt{v_3^2+\braket{H_{45}^Z}^2 }.
\end{split}
\end{equation} 
Physically, it may be interpreted as follows: inside the original superpotential $ W_Y $ lie the terms
\begin{equation}
	\mathcal{W}_Y 
	\supset \chi_3 \chi_3 H^u_{10}
	\supset \frac{v_3^2}{v_3^2 + \braket{H_{45}^Z}^2} \,
	t \, t \, H^u_{10} , 
\end{equation}
which generate renormalisable mass terms for the top quark and the third Dirac neutrino at the electroweak scale.

\section{Complete derivation of mass matrices}
\label{sec:app:fullderivation}

In this section we derive the precise forms of the Yukawa and Majorana mass matrices, and cast them in terms of a minimum number of free parameters, taking into account the vacuum alignments of the adjoint Higgs superfields as well as the induced rotation in the third family couplings due to a large $ \braket{\phi_3} $.

The renormalisable superpotential in Eq.\ref{eq:WYrenorm}, including the dominant Planck-suppressed terms in Eq.\ref{eq:WYplanck}, and writing explicitly all $ \mathcal{O}(1) $ couplings, becomes,
\begin{equation}
\begin{split}
	W_Y &= \sum_{a=1,2,3} 
		\lambda^\phi_a \psi \phi_a \overbar{\chi}_a 
		+ \lambda^\chi_a \chi_a \overbar{\chi}_a H_{45}^Z 
		+ \chi_a \chi_a \left(
			\lambda^u_a H^u_{10} 
			+ \lambda^N_a \frac{H_{\overbar{16}} H_{\overbar{16}}}{M_P}\right)
	\\ &\qquad 
	+ \sum_{b=2,3}
		\overbar{\chi}_b \chi_b^\prime \left(\lambda^{X}_b H_{45}^X + \lambda^{Y}_b H_{45}^Y\right) 
		+ \lambda^d_b \chi^\prime_b \chi^\prime_b H_{10}^d 
	\\ &\qquad 
		+ \lambda_{12}^d\chi_1\chi_2H^d_{10}
		+ \lambda^\rho_3 \rho \chi_3 H_{\overbar{16}} +M_\rho \rho\rho
		+ \lambda^d_P \frac{\psi \psi \phi_3 H^d_{10}}{M_P}.
\label{eq:WYrenormfull}
\end{split}
\end{equation}
The flavon vacuum alignments, obtained from \cite{King:2016yvg}, which preserve the generator product $ SU $, are (as in Eq.~\ref{eq:flavons})
\begin{equation}
	\braket{\phi_1} = v_1 \pmatr{1\\3\\-1}, \quad 
	\braket{\phi_2} = v_2 \pmatr{0\\1\\-1}, \quad
	\braket{\phi_3} = v_3 \pmatr{0\\1\\0},
\end{equation}
with the hierarchy $ v_1 \ll v_2 \ll v_3 $.
The singlet product which occurs in $\psi \phi_a$ above, i.e. $ 3^\prime \times 3^\prime \to 1 $, is given by
\begin{equation}
	(A B) = A_1 B_1 + A_2 B_3 + A_3 B_2.
\end{equation}
To account for this nontrivial product as well as the field redefinition $ \psi_2 \to - \psi_2 $ (this overall sign is unphysical), we define the vectors
\begin{equation}
	\braket{\tilde{\phi}_i} = I_{S_4}\braket{\phi_i},
	\quad \mathrm{with} \quad 
	I_{S_4} = \pmatr{1&0&0\\0&0&-1\\0&1&0}.
\end{equation}
In the new variables, the alignments become,
\begin{equation}
	\braket{\tilde\phi_1} = v_1 \pmatr{1\\1\\3}, \quad 
	\braket{\tilde\phi_2} = v_2 \pmatr{0\\1\\1}, \quad
	\braket{\tilde\phi_3} = v_3 \pmatr{0\\0\\1}.
\end{equation}

Next, we introduce notation to specify the relevant components of a VEV $ \braket{H_{45}^k} $, corresponding to unique CG factors. 
The index $ k $ labels the adjoint, i.e. $ k = X,Y,Z $, or $ B-L $.
$\psi$ couples to the adjoint VEVs via $ \chi $ messengers. 
After the GUT is broken and $ \psi $ is decomposed into multiplets of the SM gauge group, the part of an adjoint VEV which couples to a given multiplet $ f $ is then denoted 
\begin{equation}
	H_{45}^k \to \hvev{k}{f},
\label{eq:h45notation}
\end{equation}
where $ f = Q, u^c, d^c, L, e^c, $ or $ \nu^c $.
The $ H_{\overbar{16}} $ gets a VEV in the direction which preserves $ SU(5) $, which we call the (singlet) $ \nu^c $ direction. 
Its VEV only affects the RH neutrino mass matrix and is simply denoted $ v_{\overbar{16}} $.

We extract the Yukawa matrices from diagrams in Figs.~\ref{fig:hu}-\ref{fig:h16}. 
Taking into account nontrivial $ S_4 $ products (as above), we have
{
\newcommand{\tphi}{{\tilde{\phi}}}
\begin{align}
	Y^{u}_{ij} &=
		\sum_{a=1,2} \lambda^u_a 
		\frac{(\lambda^\phi_a)^2 \braket{\tphi_a}_i \braket{\tphi_a}_j}%
		{(\lambda^\chi_a)^2 \hvev{Z}{Q} \hvev{Z}{u^c}}+\frac{(\lambda^\phi_3)^2 \braket{\tphi_3}_i \braket{\tphi_3}_j}%
		{(\lambda^\phi_3)^2 v_3^2+(\lambda^\chi_3)^2 \hvev{Z}{Q} \hvev{Z}{u^c}}, \label{eq:tMu} \\
	Y^{\nu}_{ij} &=
		\sum_{a=1,2} \lambda^u_a 
		\frac{(\lambda^\phi_a)^2 \braket{\tphi_a}_i \braket{\tphi_a}_j}%
		{(\lambda^\chi_a)^2 \hvev{Z}{L} \hvev{Z}{\nu^c}}+\frac{(\lambda^\phi_3)^2 \braket{\tphi_3}_i \braket{\tphi_3}_j}%
		{(\lambda^\phi_3)^2 v_3^2+(\lambda^\chi_3)^2 \hvev{Z}{L} \hvev{Z}{\nu^c}}, \label{eq:tMnuD} \\
	M^{\mathrm{R}}_{ij} &=  
		\sum_{a=1,2} \frac{\lambda^N_a v_{\overbar{16}}^2}{M_P}
		\frac{(\lambda^\phi_a)^2 \braket{\tphi_a}_i \braket{\tphi_a}_j}%
		{(\lambda^\chi_a)^2 \hvev{Z}{\nu^c} \hvev{Z}{\nu^c}}
		 \nonumber 
		\\ &\qquad + v_{\overbar{16}}^2 \left(\frac{(\lambda^\rho_3)^2}{M_\rho}+\frac{\lambda^N_3}{M_P}\right)
		\frac{(\lambda^\phi_3)^2 \braket{\tphi_3}_i \braket{\tphi_3}_j}%
		{(\lambda^\phi_3)^2 v_3^2+(\lambda^\chi_3)^2 \hvev{Z}{\nu^c} \hvev{Z}{\nu^c}},
		\label{eq:tMnuM} \\
	Y^{d}_{ij} &=
		 \lambda^d_2 
		\frac{(\lambda^\phi_2)^2\braket{\tphi_2}_i\braket{\tphi_2}_j}%
		{[\lambda_2^X \braket{H_{45}^X} + \lambda_2^Y \braket{H_{45}^Y}]_Q 
		 [\lambda_2^X \braket{H_{45}^X} + \lambda_2Y \braket{H_{45}^Y}]_{d^c}} 
		\nonumber \\& \qquad
		+ \lambda^d_3 
		\frac{(\lambda^\phi_3)^2\braket{\tphi_3}_i\braket{\tphi_3}_j}%
		{(\lambda^\phi_3)^2v_3^2+[\lambda_3^X \braket{H_{45}^X} + \lambda_3^Y \braket{H_{45}^Y}]_Q 
		 [\lambda_3^X \braket{H_{45}^X} + \lambda_3Y \braket{H_{45}^Y}]_{d^c}} 
		\nonumber \\& \qquad
		+ \lambda^d_{12} 
		\frac{\lambda^\phi_1\lambda^\phi_2 \braket{\tphi_1}_i \braket{\tphi_2}_j}%
		{\lambda^\chi_1 \lambda^\chi_2 \hvev{Z}{Q} \hvev{Z}{d^c}} 
		+ \lambda^d_P \frac{ Y_P v_3}{M_P} , \label{eq:tMd} \\
	Y^{e}_{ij} &=
		 \lambda^d_2 
		\frac{(\lambda^\phi_2)^2\braket{\tphi_2}_i\braket{\tphi_2}_j}%
		{[\lambda_2^X \braket{H_{45}^X} + \lambda_2^Y \braket{H_{45}^Y}]_L 
		 [\lambda_2^X \braket{H_{45}^X} + \lambda_2Y \braket{H_{45}^Y}]_{e^c}} 
		\nonumber \\& \qquad
		+ \lambda^d_3 
		\frac{(\lambda^\phi_3)^2\braket{\tphi_3}_i\braket{\tphi_3}_j}%
		{(\lambda^\phi_3)^2v_3^2+[\lambda_3^X \braket{H_{45}^X} + \lambda_3^Y \braket{H_{45}^Y}]_L
		 [\lambda_3^X \braket{H_{45}^X} + \lambda_3Y \braket{H_{45}^Y}]_{e^c}} 
		\nonumber \\& \qquad
		+ \lambda^d_{12} 
		\frac{\lambda^\phi_1\lambda^\phi_2 \braket{\tphi_1}_i \braket{\tphi_2}_j}%
		{\lambda^\chi_1 \lambda^\chi_2 \hvev{Z}{L} \hvev{Z}{e^c}} 
		+ \lambda^d_P \frac{ Y_P v_3}{M_P} , 
		\label{eq:tMe}
\end{align}
where $v_3=|\braket{\phi_3}|$, $ v_u Y^{\nu}_{ij} $ is the Dirac neutrino mass matrix and $ M^{\mathrm{R}}_{ij} $ is the RH Majorana matrix, respectively. 
The last term in Eq.~\ref{eq:WYrenormfull} is a singlet coming from three $S_4$ triplets and gives rise to the last terms in Eqs.~\ref{eq:tMd} and \ref{eq:tMe}, respectively, where $ Y_P $ is the numerical matrix defined in Eq.~\ref{eq:Ymatrices}.
}

Finally, we take into account the effect of mixing between the state $ \psi_3 $ and messenger $ \chi_3 $. 
This mixing provides additional contributions to the fermion mass matrices in the form of coefficients multiplying the third rows and columns. 
The size of each coefficient depends on the CG factors and the ratio(s) of $ v_3 $ to adjoint Higgs VEVs $ v_{45}^k $, for $ k = X,Y,\chi $. 
In the limit where $ v_3 \ll v_{45}^k $, all these coefficients are 1, corresponding to a negligible amount of $ \chi_3 $ being mixed into the physical state. 
This is exactly what occurs for the other two families: the massless states are aligned almost exactly with the states $ \psi_{1,2} $.
Generally, any significant deviation would require a tuning among CG factors and $ \mathcal{O}(1) $ parameters $ \lambda $.  
We do not expect these factors to have a large effect on mixing, hence we set them all to one for simplicity.

\section{SUSY threshold corrections}
\label{sec:app:susy}

An analysis of the running of MSSM Yukawa parameters up to the GUT scale has been performed in \cite{Antusch:2013jca}, where they propose a useful parametrisation of $ \tan \beta $-enhanced corrections to the charged fermion Yukawa couplings and quark mixing angles.

The matching conditions at the SUSY scale $ M_\mathrm{SUSY} $ are parametrised in terms of four parameters $ \bar{\eta}_{q,b,\ell} $ and $ \bar{\beta} $, as
\begin{equation}
\begin{split}
y^\mathrm{MSSM}_{u,c,t} 	&\simeq y^\mathrm{SM}_{u,c,t} \csc \bar{\beta} , \\
y^\mathrm{MSSM}_{d,s} 		&\simeq (1 + \bar{\eta}_q)^{-1} \, y^\mathrm{SM}_{d,s} \sec \bar{\beta} , \\
y^\mathrm{MSSM}_{b} 		&\simeq (1 + \bar{\eta}_b)^{-1} \, y^\mathrm{SM}_{b} \sec \bar{\beta} , \\
y^\mathrm{MSSM}_{e,\mu} 	&\simeq (1 + \bar{\eta}_\ell)^{-1} \, y^\mathrm{SM}_{e,\mu} \sec \bar{\beta} , \\
y^\mathrm{MSSM}_{\tau} 		&\simeq y^\mathrm{MSSM}_{\tau} \sec \bar{\beta} .
\end{split}
\label{eq:MSSMyukawas}
\end{equation}
The CKM matrix also gets corrections
\begin{equation}
\begin{split}
\theta^{q,\mathrm{MSSM}}_{i3} 	&\simeq \frac{1 + \bar{\eta}_b}{1 + \bar{\eta}_q} \, \theta^{q,\mathrm{SM}}_{i3} , \\
\theta^{q,\mathrm{MSSM}}_{12} 	&\simeq \theta^{q,\mathrm{SM}}_{12} , \\
\delta^{q,\mathrm{MSSM}} 		&\simeq \delta^{q,\mathrm{SM}} .
\end{split}
\label{eq:MSSMmixingangles}
\end{equation}
To a very good approximation $\theta_{12}^q$ and $\delta^q$ are not affected by the threshold corrections.
The running of couplings $ y^{\mathrm{MSSM}}_i $ up to the GUT scale, $ y^{\mathrm{MSSM}}_i \to y^{\mathrm{MSSM@GUT}}_i $, depends to a good approximation only on $\bar{\eta}_b$ and $\tan \bar{\beta}$. 
In the limit where threshold corrections to $y_\tau$ are negligible, $\overbar{\beta}$ reduces to the usual $\beta$. 
We will assume just such a scenario. 
We will also set $ \bar{\eta}_q = \bar{\eta}_\ell = 0 $ for simplicity, as these are found not to affect the quality of the fit. 
The associated shift in down-type and charged lepton Yukawa couplings can be largely subsumed into available free parameters $ y^{d,e}_{12} $ $ y^{d,e}_2 $.
Similarly, we fix $ M_\mathrm{SUSY} = 1 $ TeV. 
Slightly larger values are allowed, up to $ \mathcal{O}(10) $ TeV, but the effect on the fit is minor.
Conversely, the remaining SUSY parameter, $ \bar{\eta}_b $, will be important and prefers a large (negative) value. 
A very good fit is found for $ \bar{\eta}_b = -0.8 $, for which results are presented in this paper.
Meanwhile, the neutrino masses and mixing angles are expected to be largely insensitive to group running.

\end{document}